\documentclass[10pt, conference, letterpaper]{IEEEtran}
\IEEEoverridecommandlockouts
\usepackage{amsmath,amssymb,amsfonts}
\usepackage{algorithmic}
\usepackage{graphicx}
\usepackage{subfig}
\usepackage{textcomp}
\usepackage[dvipsnames, svgnames, x11names]{xcolor}
\usepackage{multirow}
\usepackage{booktabs}
\usepackage[ruled,vlined,linesnumbered]{algorithm2e}
\usepackage[numbers,sort&compress]{natbib}
\usepackage{hyperref}
\usepackage{amsthm}
\usepackage{bm}
\usepackage{threeparttable}

\newtheorem{assumption}{Assumption}
\newtheorem{theorem}{Theorem}

\newtheorem{definition}{Definition}

\newcommand{\noaistats}[1]{}
\newcommand{\SUB}[1]{\ENSURE \hspace{-0.3in} \textbf{#1}}

\def\BibTeX{{\rm B\kern-.05em{\sc i\kern-.025em b}\kern-.08em
		T\kern-.1667em\lower.7ex\hbox{E}\kern-.125emX}}

\begin{document}
	\title{FedBIAD: Communication-Efficient and Accuracy-Guaranteed Federated Learning with Bayesian Inference-Based Adaptive Dropout}
	
	\author{\IEEEauthorblockN{Jingjing Xue\IEEEauthorrefmark{1}\IEEEauthorrefmark{2}, Min Liu\IEEEauthorrefmark{1}\IEEEauthorrefmark{2}\IEEEauthorrefmark{3}, Sheng Sun\IEEEauthorrefmark{1}, Yuwei Wang\IEEEauthorrefmark{1}, Hui Jiang\IEEEauthorrefmark{1}\IEEEauthorrefmark{2}, and Xuefeng Jiang\IEEEauthorrefmark{1}\IEEEauthorrefmark{2}}
	\IEEEauthorblockA{\IEEEauthorrefmark{1}Institute of Computing Technology, Chinese Academy of Sciences, Beijing, China}
	\IEEEauthorblockA{\IEEEauthorrefmark{2}University of Chinese Academy of Sciences, Beijing, China}
	\IEEEauthorblockA{Email: \{xuejingjing20g, liumin, sunsheng, ywwang, jianghui, jiangxuefeng21b\}@ict.ac.cn}
	}
	
	\maketitle

	\begin{abstract}
		Federated Learning (FL) emerges as a distributed machine learning paradigm without end-user data transmission, effectively avoiding privacy leakage. Participating devices in FL are usually bandwidth-constrained, and the uplink is much slower than the downlink in wireless networks, which causes a severe uplink communication bottleneck. A prominent direction to alleviate this problem is federated dropout, which drops fractional weights of local models. However, existing federated dropout studies focus on random or ordered dropout and lack theoretical support, resulting in unguaranteed performance. In this paper, we propose Federated learning with Bayesian Inference-based Adaptive Dropout (FedBIAD), which regards weight rows of local models as probability distributions and adaptively drops partial weight rows based on importance indicators correlated with the trend of local training loss. By applying FedBIAD, each client adaptively selects a high-quality dropping pattern with accurate approximations and only transmits parameters of non-dropped weight rows to mitigate uplink costs while improving accuracy. Theoretical analysis demonstrates that the convergence rate of the average generalization error of FedBIAD is minimax optimal up to a squared logarithmic factor. Extensive experiments on image classification and next-word prediction show that compared with status quo approaches, FedBIAD provides 2$\times$ uplink reduction with an accuracy increase of up to 2.41\% even on non-Independent and Identically Distributed (non-IID) data, which brings up to 72\% decrease in training time.
	\end{abstract}
	
	\begin{IEEEkeywords}
		federated learning, communication bottleneck, adaptive dropout, Bayesian inference
	\end{IEEEkeywords}
	
	\section{Introduction}
	
	The increased prevalence of smart end-user devices has led to the significant growth of the data generated by edge networks. To utilize such decentralized data without possible leakage of end-user privacy,
	Federated Learning (FL) \cite{fedavg_2017, FL_keyboard_pre_2018} has emerged as a distributed paradigm for training a global model over massive devices without raw data transmission. In particular, participating end devices (i.e., clients) train locally on their own data and periodically upload local model updates to a central server for global aggregation \cite{FL_introduce_2022}.
	
	The clients in FL are usually limited by lower bandwidth in wireless networks, whereas millions of model parameters have to be uploaded. For example, a large model with 488.08 MegaByte (MB) weights is trained for speech recognition \cite{dgc_2018}. After hundreds of iterations, the total communication of every client can grow to more than a PetaByte (PB) \cite{STC_2020}. The mismatch between communication resources and transmission demand constitutes the communication bottleneck of FL. As the uplink is typically much slower than the downlink, e.g., the T-Mobile 5G wireless network at 110.6 Mbps down vs. 14.0Mbps up \cite{down_up}, it is indispensable to design communication-efficient FL schemes for mitigating uplink overhead.
	
	
	
	\begin{figure}
		\centering
		{\includegraphics[width=0.49\textwidth]{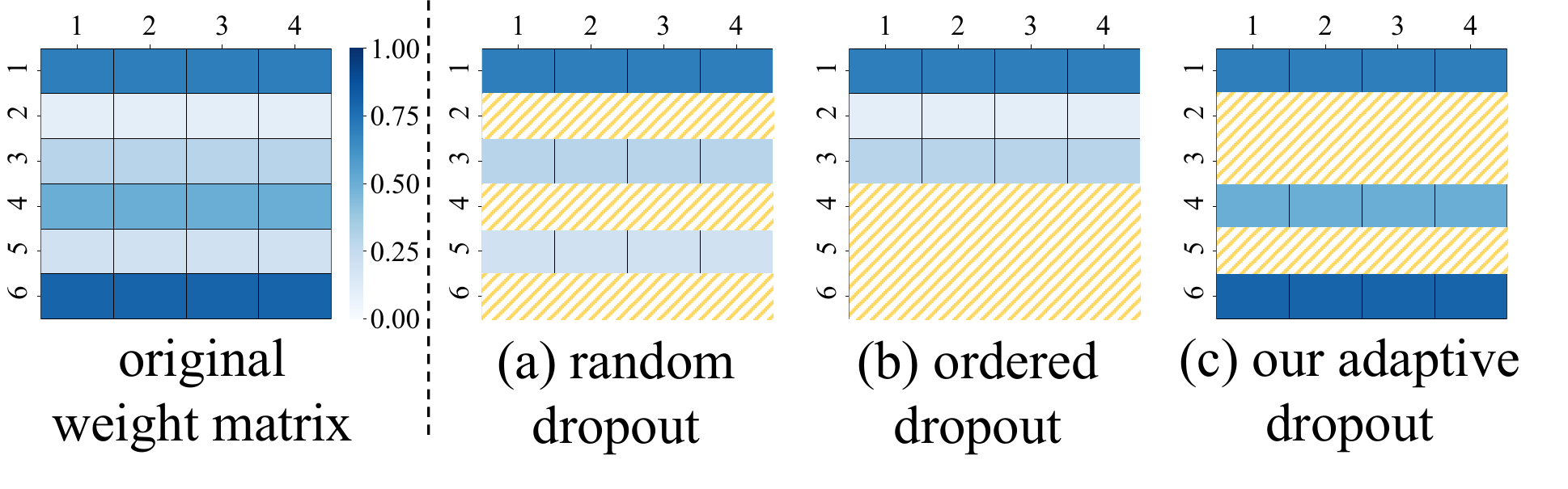}}
		\caption{Comparison of dropout, where blue shading is importance of weight rows, and yellow masks denote dropped rows.}
		\label{dropout_method}
		\vskip -0.2in
	\end{figure}
	
	Several pioneering studies have attempted to alleviate the uplink burden \cite{Lossy_compress_2016, CMFL_2019, FedPAQ_2020, SmartIdx_2022}. A feasible solution is to transmit compressed parameters \emph{after} local training (i.e., sketched compression \cite{Lossy_compress_2016}), which learns a \emph{full} model update and compresses it. Yet this method inevitably introduces the noise to models, and the noise is accumulated over long-term learning, leading to accuracy degradation \cite{Lossy_compress_2016, signSGD_2018}. Another solution is model compression \emph{in} local training, where clients train on local \emph{shrunk} models. It is flexible for clients to adjust compressed model structures based on local data. Federated Dropout (FD) \cite{fed_dropout_2019,federated_dropout_2021,fjord_2021,AFD_2021} is a prominent approach for model compression in local training, which dynamically drops fractional neurons for each client. It implies that fractional rows or columns of weight matrices are zeroed and not uploaded to the server.

\begin{figure}
		\centering
		{\includegraphics[width=0.45\textwidth]{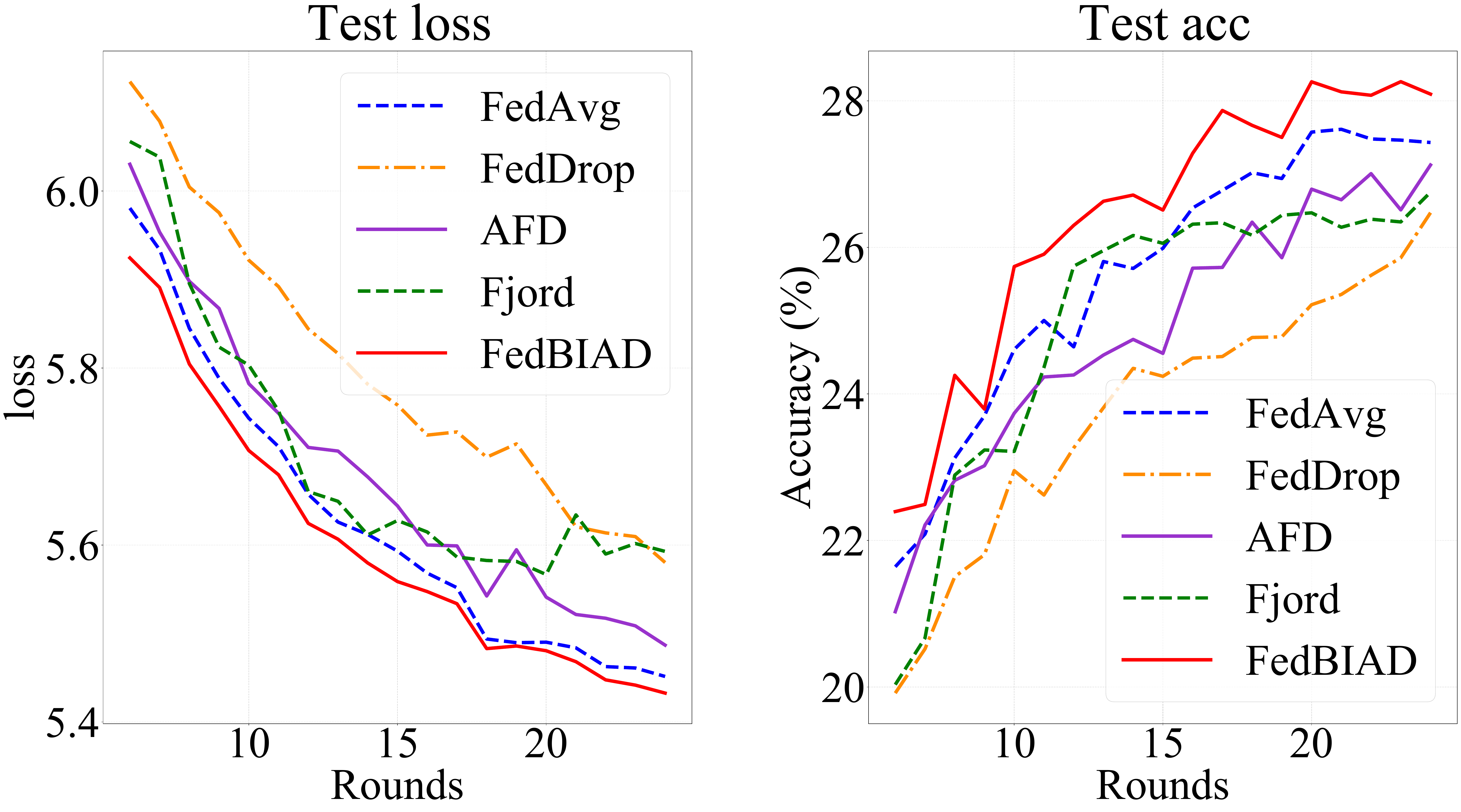}}
		\caption{Comparison of test results for different methods.}
		\label{dropout_rnn_compare}
		\vskip -0.15in
	\end{figure}

	However, existing FD works still face two challenges: unguaranteed accuracy and lack of theoretical analysis. Firstly, current mainstream FD researches concentrate on random dropout (such as FedDrop \cite{fed_dropout_2019, federated_dropout_2021}) or ordered dropout (such as Fjord \cite{fjord_2021}, which drops partial back rows in weight matrices). They ignore the different importance of weights, meaning that some significant weights are dropped (see Fig. \ref{dropout_method}) at the expense of accuracy. AFD \cite{AFD_2021} advances federated dropout by designing score maps in the server to determine dropping structures, while clients cannot adjust dropping structures during local training, leading to less flexibility for dropout. In addition, FedDrop and AFD are limited in the application scope. They have not extended to recurrent connections of Recurrent Neural Networks (RNN), where the deviation introduced by dropout is amplified in long sequences \cite{variational_rnn_2016}, resulting in poor performance \cite{bayer_fast_2014}. Empirically, we execute next-word prediction using a Long Short-Term Memory (LSTM) model on the PTB dataset \cite{ptb_1993}. As shown in Fig. \ref{dropout_rnn_compare}, FedDrop, AFD, and Fjord all reveal worse performance than a basic FL algorithm, FedAvg. Furthermore, there is no theoretical explanation provided in existing FD works. It is not convincing without error bound and convergence proof. Thus, there is a need for a novel dropout strategy with guaranteed accuracy and theoretical interpretation in federated settings.
	
	Motivated by research at the intersection of Bayesian inference and deep learning \cite{Bayesian_dropout_2016}, the dropout can be interpreted by the spike-and-slab distribution-based Bayesian inference, where each row of weight matrices in the Deep Neural Network (DNN) follows an independent spike-and-slab prior \cite{spike_prior_2014}. As stated in \cite{Convergence_Bayes_2020}, spike-and-slab distributions zero out partial weight rows and sample the remaining rows from Gaussian \cite{spike_with_Guassian_2016, bayesian_sparse_2017}, equivalent to dropouts of corresponding activations in DNN. However, existing Bayesian interpretations of dropout \cite{variational_rnn_2016, Convergence_Bayes_2020} focus on centralized training. For boosting federated dropout, it is crucial to embrace spike-and-slab distribution-based Bayesian inference in distributed FL scenarios.

	In light of the above observations, we propose Federated learning with Bayesian Inference-based Adaptive Dropout (FedBIAD), which enables uplink overhead mitigation while improving performance and guaranteeing convergence. Specifically, we introduce Bayesian inference with spike-and-slab prior into local models, which infuses probabilistic distributions into models and is robust to overfitting on limited client data. On this basis, FedBIAD allows clients to adopt adaptive dropout based on importance indicators that evaluate the effect of each weight row on training loss, aiming at exploring a high-quality dropout for each client to enhance accuracy. Consequently, each client adaptively picks the sub-model structure with more accurate approximations and only transfers the corresponding parameters of the sub-model. Moreover, the extension of Bayesian inference provides a novel insight into the application of federated dropout to recurrent connections, and FedBIAD can empirically combine with existing sketched compressions, further alleviating the communication burden.
	
	Our main contributions can be summarized as follows:
	\begin{itemize}
		\item[$\bullet$] Considering constrained uplink communication resources and unguaranteed performance, we propose a novel FL algorithm, FedBIAD, which allows clients to adaptively drop partial weight rows based on the trend of training loss. In FedBIAD, we additionally devise an importance indicator that measures the importance of weight rows for each client. This indicator contributes to looking for a high-quality dropout with performance improvement.
		\item[$\bullet$]  To obtain theoretical support, we perform a formal analysis of the generalization error bound and convergence property. Theoretical results demonstrate that the convergence rate of the average generalization error of FedBIAD is minimax optimal up to a squared logarithmic factor.
		\item[$\bullet$]  We conduct extensive experiments on image classification and next-word prediction. The results show that FedBIAD outperforms baselines in uplink costs and test accuracy even on non-IID datasets. Compared to status quo approaches, FedBIAD provides 2$\times$ reduction in uplink costs with an accuracy increase of up to 2.41\%, further shortening the training time by up to 72\%. We empirically combine FedBIAD with a sketched compression, DGC \cite{dgc_2018}, which reduces up to 2$\times$ uplink costs while improving accuracy by up to 2.26\% compared with naive DGC.
	\end{itemize}
	
	\section{Related Work}
	
	The communication bottleneck of FL has attracted tremendous attention, and some efforts have been devoted to solving this challenge, which is roughly divided into two directions: decreasing the total communication rounds and reducing transmitted parameters per round. 
	
	\textbf{Decreasing communication rounds.} FedAvg \cite{fedavg_2017} performs multiple local training iterations in a round to reduce communication frequency. \citet{Dynamic_BatcSize_2019} find that the dynamic increase of batch size is beneficial for fast convergence to reduce communication rounds. \citet{pretraing_FL_2020} propose that pretraining models could speed up convergence and lessen rounds. However, these approaches cannot avoid the simultaneous transmission of a huge number of parameters in the wireless communication link for each round, leading to network exhaustion and transmission deceleration.
	
	\textbf{Reducing transmitted parameters.} There are mainly two strategies. The first type is reducing the number of clients that upload parameters to the server. For example, CMFL \cite{CMFL_2019} excludes the clients with irrelevant updates from parameter uploading, and FedMed \cite{fedmed_2020} considers that only clients with lower training loss are allowed to transmit parameters. Although the above methods mitigate uplink costs, they waste computation resources of devices that train locally but do not participate in global aggregation. The second type is reducing the transmitted parameters of each client. SignSGD \cite{signSGD_2018} and FedPAQ \cite{FedPAQ_2020} quantize weights into low-precision values to reduce communication volume. DGC \cite{dgc_2018} and SmartIdx \cite{SmartIdx_2022} propose to transfer fractional gradients with higher magnitudes. These sketched approaches compress weights after local training, inevitably pulling noise into models. The noise is accumulated in the long training, adversely affecting model accuracy. Besides, low rank \cite{Lossy_compress_2016} enforces every weight matrix as the product of two low-rank matrices, one of which is fixed and the other is trained and transmitted. This fixed structure restricts model updates \cite{AFD_2021}, resulting in poor performance \cite{Lossy_compress_2016}.
	
	Instead of the fixed structure, federated dropout \cite{fed_dropout_2019, federated_dropout_2021} dynamically selects smaller models by the dropout of neurons to alleviate the communication burden. \citet{fed_dropout_2019} first extend dropout to FL, which randomly drops a fixed number of neurons for each participant client. This random dropout ignores the importance of weights, so some significant weights are discarded, resulting in the sacrifice of accuracy. More recently, FjORD \cite{fjord_2021} proposes ordered dropout that directly drops adjacent neurons of models. It assumes that partial back rows in the weight matrix of each model layer have less effect on accuracy, which can be dropped. However, this ordered theory has only been proved in linear mapping and has not been derived in non-linear DNN. Besides, FedMP \cite{FedMP_2022} directly prunes weights with lower magnitudes in clients without considering their effect on training loss. These methods also cannot achieve satisfying performance. Furthermore, although existing works have been evaluated through extensive experiments, there is no theoretical analysis to guarantee reliability. Without error bound and convergence proof, we cannot trust them to achieve consistent performance in FL.
	
	Different from the above methods, this work studies federated learning with adaptive dropout, where each client adaptively drops weight rows based on the importance indicator correlated with the local loss changes. Furthermore, considering the generalization properties of Bayesian inference, the convergence of average generalization error can be proven.
	
	\section{Bayesian Interpretation of Dropout}
	
	\subsection{Formulations for Deep Neural Network}
	\label{formulations}
	
	Deep Learning (DL) constructs an input-output mapping through a Deep Neural Network (DNN). Given a dataset $(X, Y)$, for any $(\bm{x},\,y) \in (X,\,Y)$, there is an input $\bm{x}$ and an unknown true function $f_0$ such that $y = f_0(\bm{x})$. The goal of DL is to learn a DNN model $f_{\theta}$ such that $f_{\theta}(\bm{x}) \thickapprox f_0(\bm{x})$, where $\theta$ denotes the set of weights in the model. 
	
	We consider two types of commonly used DNNs: the non-recurrent neural network and the Recurrent Neural Network (RNN). Starting by a non-recurrent neural network with $L$ layers, $f_\theta: \mathbb{R}^d \to \mathbb{R}$ is written recursively as:
	\begin{equation*}
		x_l =\rho(W_l x_{l-1}), \;\, f_\theta(\bm{x}) = x_L \quad \text{for}\;\; l = 1, \ldots,  L
	\end{equation*}
	where $x_0 = \bm{x}$, $\rho$ is an activation function, and $W_l$ is the weight matrix in the $l$-th layer of DNN such that $\theta = \{W_1, \dots, W_L\}$. The input dimension is $d$. For ease of analysis, we use an equal-width DNN as \cite{Convergence_Bayes_2020} and \cite{pFedBayes_2022}, i,e., the number of units in each hidden layer is equal to $D$, where $d \leq D$. 
	
	For sequence learning tasks, such as time series forecasting and next-word prediction, the RNN with memory correlation ability is more effective. In this case, the input is a sequence of feature vectors $[x_1, \ldots, x_L], x_l\in \mathbb{R}^d$, and $f_\theta$ can be expressed in a different form. The input sequence is mapped by 
	$$
	h_l =\varrho(W_xx_l + W_hh_{l-1}),\;\, f_\theta(\bm{x})= \phi(h_L) \quad \text{for}\; l =  1, \ldots,  L
	$$
	with $h_0=\overrightarrow{0}$. $W_x\in \mathbb{R}^{D\times d}$ is the input-hidden weight matrix, and $W_h \in \mathbb{R}^{D \times D}$ is the hidden-hidden weight matrix (denoted recurrent connections of RNN) such that $\theta = \{W_x, W_h\}$. Besides, $\varrho$ and $\phi$ are activation functions.
	
	We set $N$ as the total number of weights in $\theta$ and $\mathrm{J}$ as the number of rows in all weight matrices. The $\textbf{w}_{j}$ denotes the $j$-th row of weight matrices. The unsparse number $S$ is the number of nonzero weights in $\theta$. In this way, the model architecture is determined by $(S,L,D)$. $\Theta_{S,L,D}$ represents the solution space composed of all feasible weights.
	
	\subsection{Variational Bayesian Inference}
	\label{variational_bayesian_inference}
	
	From the Bayesian perspective, we consider $\theta$ as random variables following prior $\pi$. Given the dataset ($X, Y$), we aim to find out the posterior $\pi(\theta|X,Y)$, which is usually intractable because the posterior is high-dimensional and non-convex in DNN \cite{Bayes_posteriors_2021}. Variational Inference (VI) \cite{varia_inference_2011} is one of the most efficient solutions for the posterior studied in the last decade \cite{bayes_mcmc_2011, varitional_cnn_2017, bayes_mc_2021}, which searches for a variational approximation to estimate $\pi(\theta|X,Y)$. Based on VI, \cite{tempered_posteriors_2019} considers that the likelihood $p(y|\bm{x},\theta)$ could be replaced by $\big[ p(y|\bm{x},\theta) \big]^\alpha$ for any $\alpha \in (0,1)$, leading to the tempered posterior 
	\begin{equation}
		\pi_{m,\alpha} \propto \sum^m_{i=1} \alpha \ln p(y_i|\bm{x}_i,\theta) \pi(\mathrm{d}\theta),
	\end{equation}
	where $m$ is the number of samples. The proof of concentration of the tempered posterior $\pi_{m,\alpha}$ needs fewer assumptions than $\pi(\theta|X,Y)$ \cite{tempered_posteriors_2019}. Following it, we pay attention to a variational approximation $\Tilde{\pi}$ of the tempered posterior $\pi_{m,\alpha}$. 
	
	For any $(\bm{x}_i,y_i) \in (X,Y),\, i\in\{1,\ldots,m\}$, the $p(y_i|\bm{x}_i,\theta)$ follows a Gaussian with a likelihood variance $\sigma^2$. Our objective is to explore a variational approximation $\Tilde{\pi}$ to estimate the tempered posterior $\pi_{m,\alpha}$ for minimizing the loss function
	\begin{equation}
		\label{loss}
		\mathcal{L}(\Tilde{\pi}) = \frac{\alpha}{2\sigma^2} \sum_{i=1}^{m} \int \big( y_i - f_{\theta}(\bm{x}_i))^2 \Tilde{\pi}(\mathrm{d}\theta \big) + KL( \Tilde{\pi}\|\pi).
	\end{equation}
	The second item from the right in (\ref{loss}) has been proven to approximate L2 regularisation \cite{variational_rnn_2016, spike_with_Guassian_2016}. Hence, variational Bayesian inference avoids overfitting and easily learns from small samples. Inspired by \cite{Convergence_Bayes_2020}, Definition \ref{def:def1} gives an expression for the generalization error of variational approximation.
	
	\begin{definition}
		\label{def:def1}
		The generalization error of the variational approximation $\Tilde{\pi}$ is
		$G_e(\Tilde{\pi}) = \mathbb{E} \Big[ \int \|{f_{\theta}-f_0}\|^2_2 \Tilde{\pi}(\mathrm{d}\theta) \Big]$.
	\end{definition}
	
	\subsection{Bayesian Inference-based Dropout}
	
	The Bayesian inference with spike-and-slab distribution can act as a relevant proxy of the dropout \cite{Convergence_Bayes_2020}. As mentioned in \cite{federated_dropout_2021}, the dropout technique zeros out a fixed number of weights, while the idea of setting a subset of weights to zero can be realized by placing spike-and-slab distributions over weights \cite{spike_with_Guassian_2016}. Motivated by \cite{variational_rnn_2016} and \cite{spike_with_Guassian_2016}, we define the spike-and-slab distributions to factorize over the weight matrices, where each weight row $\textbf{w}_{j}$ follows
	\begin{equation}
		\tilde{\pi}(\textbf{w}_j) = p \delta(0)
		+(1-p)\mathcal{N}(\bm{\mu}_j, \Tilde{s}^2I),
	\end{equation}
	with dropout rate $p$ given in advance, the variational parameter (row vector) $\bm{\mu}_j$ learned in the local training, and posterior variance $\Tilde{s}^2$. Here, $\delta(0)$ is an impulse function. We suppose that posterior variance $\Tilde{s}^2$ is constant for any weight rows, which is common in Bayesian variational inference, as referred to \cite{variational_rnn_2016, Convergence_Bayes_2020}. The precise setting of the posterior variance $\Tilde{s}^2$ for FedBIAD will be detailed in Section \ref{convergence_analysis}. Based on the setting of constant posterior variance, we conduct extensive experiments, and the results are shown in Section \ref{experimental_results}.
	
	For better characterizing the spike-and-slab distributions of weight rows, we define a binary vector $\bm{\beta}=[\beta_1,\ldots,\beta_\mathrm{J}]^\top \in \{0,1\}^\mathrm{J}$ as the dropping pattern to represent whether weight rows are dropped. If the dropping label $\beta_j$ of the $j$-th row is zero, then $\textbf{w}_j$ is dropped; otherwise, $\textbf{w}_j$ is held for training. Given dropout rate $p$, we compute the unsparse number $S=(1-p)\mathrm{J}\times D$ and determine the set of all possible dropping patterns $\mathcal{Z}^S_N$. Sampling $\bm{\beta}$ from $\mathcal{Z}^S_N$, the spike-and-slab distribution of $\textbf{w}_j$ is rewritten as
	\begin{equation}
		\tilde{\pi}(\textbf{w}_j) = \beta_j \mathcal{N}(\bm{\mu}_j, \Tilde{s}^2I)+(1-\beta_j)\delta(0),
	\end{equation}
	where $j\in\{1,\ldots, \mathrm{J}\}$. This work looks for  high-quality dropping patterns based on importance indicators of weight rows to reduce uplink costs while achieving excellent performance.
	
	\section{Federated Learning with Bayesian Inference-Based Adaptive Dropout (FedBIAD)}
	
	In this section, we first provide the problem formulation. Then, we elaborate on the details of FedBIAD, where each client adaptively drops fractional weight rows based on training loss changes and maintains an importance indicator, aiming at finding an effective dropping pattern to decrease loss. Moreover, we theoretically analyze the convergence of the average generalization error of FedBIAD.
	
	\subsection{Problem  Formulation}
	\label{problem_formulatiom}
	
	Given $K$ devices formed a set $\mathcal{K}$, each device $k\in\mathcal{K}$ owns its data $\mathcal{D}^k$ and true function $f_0^k$ such that $y^k=f^k_0(\bm{x}^k)$ for any $(\bm{x}^k, y^k)\in \mathcal{D}^k$. We aim to look for a global variational approximation $\tilde{\pi}^g = \mathcal{N}(U, \Tilde{s}^2I)$ from a variational set \cite{Convergence_Bayes_2020} $\mathcal{F}_{S,L,D} \subseteq \Theta_{S,L,D}$ to minimize the average loss over all clients:
	\begin{equation}
		\label{loss_goal}
		\mathcal{L}^ \ast = \min_{\tilde{\pi}^g\in\mathcal{F}_{S,L,D}}\bigg\{\frac{1}{K}\sum_{k=1}^K \frac{\mathcal{L}^k(\tilde{\pi}^g)}{|\mathcal{D}^k|}\bigg\}.
	\end{equation}
	Here, $\mathcal{L}^k$ denotes the loss of the $k$-th client over $\mathcal{D}^k$ that can be calculated by (\ref{loss}), in which $m$ is replaced by $|\mathcal{D}^k|$.
	
	Instead of using weights, we learn the variational approximation distribution in optimization problem (\ref{loss_goal}) of average loss minimization. Each client $k$ owns its local variational approximation $\tilde{\pi}^k$, which is determined by the dropping pattern $\bm{\beta}^k$, local variational parameters $U^k=[\bm{\mu}^k_1,\ldots,\bm{\mu}^k_\mathrm{J}]^\top$, and the posterior variance $\Tilde{s}^2$. As stated above, we assume that $\Tilde{s}^2$ is a constant, so client $k$ learns $U^k$ and explores $\bm{\beta}^k$ during local training. In this setting, client $k$ needs to transmit nonzero elements of $\bm{\beta}^k \circ U^k$ to the server, where we define
	\begin{equation}
		\bm{\beta}^k \circ U^k = [\beta^k_1\bm{\mu}^k_1,\, \ldots,\, \beta^k_\mathrm{J}\bm{\mu}^k_\mathrm{J}]^\top.
	\end{equation}
	Concretely, for any $j\in\{1,\ldots, \mathrm{J}\}$, the variational parameter $\bm{\mu}^k_j$ will be transmitted only if $\beta^k_j\neq 0$.
	
	\begin{figure}
		\centering
		{\includegraphics[width=0.41\textwidth]{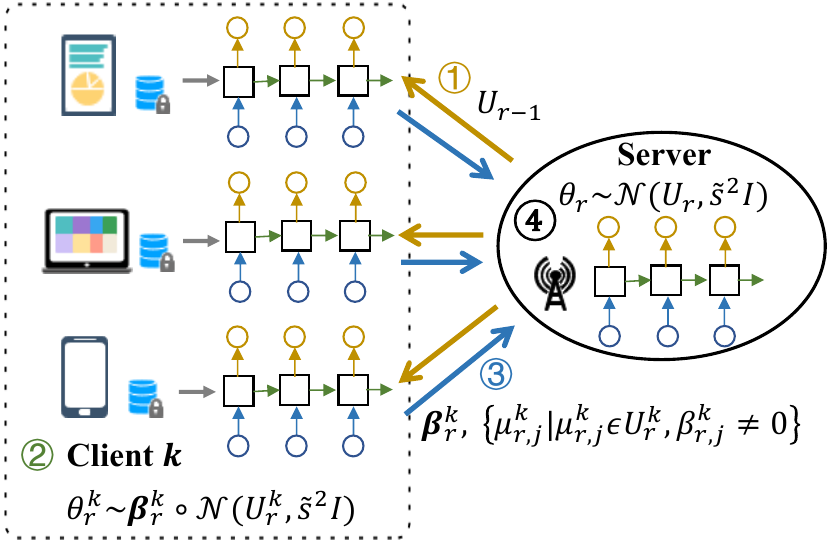}}
		\caption{An overview of FedBIAD.}
		\label{FL_frame}
		\vskip -0.1in
	\end{figure}
	
	\subsection{Overview of FedBIAD}
	
	As shown in Fig. \ref{FL_frame}, the procedure of FedBIAD in global round $r$ is as follows:
	\begin{itemize}
		\item[$\bullet$] Step 1: The server selects the client set $\mathcal{C}_r$ and sends global variational parameters $U_{r-1}$ to each client $k \in \mathcal{C}_r$.
		\item[$\bullet$] Step 2: For each client $k\in\mathcal{C}_r$, the local model is initialized by $\theta^{k,0}_r \sim \mathcal{N}(U_{r-1}, \Tilde{s}^2I)$. Then, fractional weight rows of the local model are dropped according to the dropping pattern $\bm{\beta}^{k}_r$, and the remaining sparse model is updated on local data. During local iterative training, the dropping pattern is adaptively adjusted based on changing trends of training loss. The experience of dropout is recorded in the weight score vector to characterize the importance of weight rows, which contributes to looking for a powerful dropping pattern for the client.
		\item[$\bullet$] Step 3: After $V$ training iterations, client $k \in \mathcal{C}_r$ uploads nonzero elements of variational parameters and binary dropping pattern (much less than variational parameters and can even be ignored in uplink costs) to the server.
		\item[$\bullet$] Step 4: The server reconstructs complete variational parameters based on the received binary dropping pattern for client $k \in \mathcal{C}_r$, and performs global aggregation to calculate new global variational parameters $U_{r}$ (\S\ref{aggregation}).
	\end{itemize}
	
	Specifically, FedBIAD is divided into two stages by a preset boundary $R_b$ of the global round. The difference between the two stages is the determination schemes of dropping patterns. In stage one (i.e., global round $r\leq R_b$), each client randomly samples the initial dropping pattern from $\mathcal{Z}^S_N$, and then adaptively adjusts the dropping pattern based on the trend of training loss (\S\ref{stage one}) to find a better dropping pattern during training. If global round $r>R_b$, FedBIAD enters stage two, where the dropping pattern is determined by the experience-based importance indicator (\S\ref{importance}).
	
	\subsection{Adaptive Dropout Correlated to Loss Trend}
	\label{stage one}
	
	Our main idea is to explore the dropping pattern based on the trend of local training loss. As different weight rows have different effects on loss, we should drop the rows that cannot facilitate loss reduction. Firstly, we preset the dropout rate $p$, which determines the unsparse number $S =(1-p)\mathrm{J}\times D$. Given $S$, the set $\mathcal{Z}^S_N$ is specific. The $\bm{\beta}^{k,0}_r=[\beta^{k,0}_{r,1}, \ldots, \beta^{k,0}_{r,\mathrm{J}}]^\top$ is defined as an initial dropping pattern of client $k$ in round $r$.  
	
	After receiving global variational parameters $U_{r-1}$, client $k$ initializes the local model $\theta^{k,0}_r \sim \mathcal{N}(U_{r-1}, \Tilde{s}^2I)$, randomly samples the dropping pattern $\bm{\beta}^{k,0}_r$ from $\mathcal{Z}^S_N$, and then zeros out each weight row $\textbf{w}_j$  whose dropping label $\beta^{k,0}_{r,j}=0$ in local model $\theta^{k,0}_r$. Subsequently, the remaining sparse model is trained locally. For the $v$-th iteration, the dropping pattern is denoted as $\bm{\beta}^{k,v}_r = [\beta^{k,v}_{r,1}, \ldots, \beta^{k,v}_{r,\mathrm{J}}]^\top,\, v \in \{1, \ldots, V\}$, and the variational parameters $U^{k,v}_r=[\bm{\mu}_{r,1}^{k,v},\ldots, \bm{\mu}_{r,\mathrm{J}}^{k,v}]^\top$ update by
	\begin{equation}
		\label{local_update}
		U^{k,v+1}_{r} =  U^{k,v}_r - \eta \big[\bm{\beta}^{k,v}_r \circ \nabla_U \mathcal{L}^k\big( \Tilde{\pi}^{k,v}_r(\bm{\beta}^{k,v}_r , U^{k,v}_r) \big)\big],
	\end{equation}
	where $\eta$ is the learning rate and $\Tilde{\pi}^{k,v}_r = \bm{\beta}^{k,v}_r \circ \mathcal{N}(U^{k,v}_r, \Tilde{s}^2I)$.
	
	\begin{figure}
		\begin{center}
			\centerline{\includegraphics[width=0.99\columnwidth]{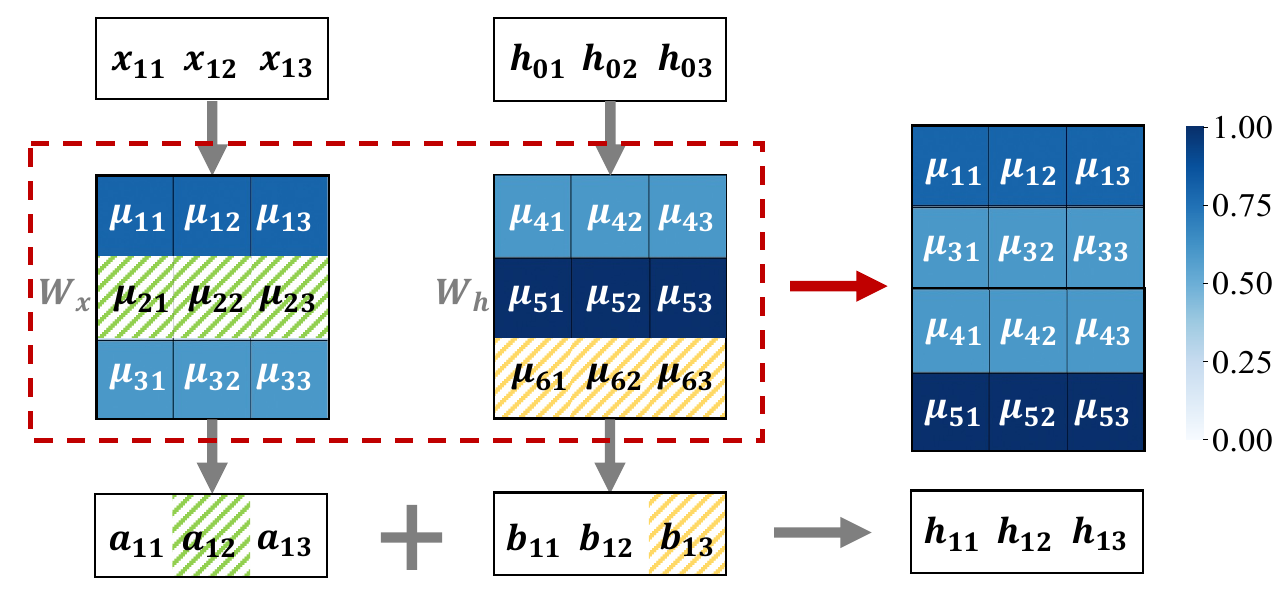}}
			\caption{The dropout based on the importance indicator. The blue shades are the normalized importance of weight rows. The green and yellow masks indicate that they are zeroed.}
			\label{weight_dropout}
		\end{center}
		\vskip -0.2in
	\end{figure}
	
	During training, client $k$ calculates the loss gap between adjacent $\tau$ iterations, described as:
	\begin{equation}
		\label{loss_diff}
		\Delta\mathcal{L}^{k,v}_r =\bar{\mathcal{L}}^{k,v}_r - \bar{\mathcal{L}}^{k, v-\tau}_r,\,\,\,v \geq 2\tau,
	\end{equation} 
	where $\bar{\mathcal{L}}^{k,v}_r = \sum_{i=v-\tau+1}^{v}{\mathcal{L}^k(\Tilde{\pi}^{k,i}_r)}$ is the average loss of the $(v-\tau+1)$-th iteration to the $v$-th iteration in the global round $r$. If $\Delta\mathcal{L}^{k,v}_r \leq 0$, the current dropping pattern $\bm{\beta}^{k,v}_r$ is favorable for loss decrease, which is used for the next $\tau$ iterations; otherwise, client $k$ resamples $\bm{\beta}^{k,v+1}_r=\ldots= \bm{\beta}^{k,v+\tau}_r$ from $\mathcal{Z}^S_N$ for training in the next $\tau$ iterations. After $V$ iterations, client $k$ drops zero row vectors, and transfers the variational parameters of the remaining weight rows to the server. Through dropout, clients can reduce transmitted parameter size so the uplink overhead can be alleviated. Note that, the amount of reduction in uplink costs is determined by the preset dropout rate $p$.
	
	\textbf{An intuitive dropout example for CNNs.} In the above description, we elaborated on the dropout for fully connected neural networks and RNNs, where the weight matrix for each model layer is two-dimensional, and we can view weights by rows \cite{variational_rnn_2016, spike_with_Guassian_2016}. Extending to Convolutional Neural Networks (CNNs), we view weights by filters and dropout is filter-wise, as referred to \cite{fed_dropout_2019, FedMP_2022}. Each client $k$ possesses a filter-wise dropping pattern $\bm{\beta}_r^{k,v}=[\beta^{k,v}_{r,1},\ldots,\beta^{k,v}_{r,J}]^\top$ for the local CNN in round $r$. For each convolutional layer in CNN, if the $j$-th filter has the dropping label $\beta^{k,v}_{r,j}=0$, all weights in this filter are zeroed out, meaning that the $j$-th filter is dropped. During local training, each client adaptively adjusts the filter-wise dropping pattern according to the loss gap calculated by (\ref{loss_diff}), as mentioned above. After that, the dropped filters in the local CNN will not be transmitted to the server.
	
	\begin{algorithm}[t]
		\begin{algorithmic}[1]
			\REQUIRE dropout rate $p$, client selection fraction $\kappa$, the  \\ number of local
			iterations $V$, boundary of stages $R_b$.
			\SUB{Initialize:} global variational parameters $U_0$, \\
			the number of selected clients $c \leftarrow \max(\lfloor \kappa \cdot K \rfloor ,\, 1)$, \\ 
			the dropping pattern set $\mathcal{Z}^S_N$, posterior variance $\Tilde{s}^2$
			\SUB{Aggregate:} \qquad \qquad \qquad \qquad \qquad \emph{\color{blue} // Run on the server}
			\FOR{each round $r = 1, 2, \dots, R$}
			\STATE $\mathcal{C}_r \leftarrow$ (random set of $c$ clients)
			\STATE Send global $U_{r-1}$ to client $k \in \mathcal{C}_r$
			\FOR{each client $k \in \mathcal{C}_r$ \textbf{in parallel}}
			\STATE $ \bm{\beta}^{k,V}_r$, $\{\bm{\mu}_{r,j}^{k,V}\} \leftarrow$ \textbf{ClientUpdate}($U_{r-1}$, $r$)
			\ENDFOR
			\STATE Reconstruct $\bm{\beta}^{k,V}_r \circ U_{r}^{k,V}$ and calculate $U_{r}$ by (\ref{agg})
			\ENDFOR
			\SUB{ClientUpdate ($U_{r-1}$, $r$):} \qquad \qquad \quad \emph{\color{blue}// Done by clients}
			\STATE Initialize $U^{k,0}_r \leftarrow U_{r-1}$ and $\theta^{k,0}_r \sim \mathcal{N}(U^{k,0}_r, \Tilde{s}^2I)$
			\IF{$r \leq R_b $}
			\STATE $\bm{\beta}^{k,0}_r= \ldots = \bm{\beta}^{k,2\tau}_r \leftarrow$ Randomly sample from \\ $\mathcal{Z}^S_N$ \qquad \qquad \qquad \qquad \qquad \qquad \quad \emph{\color{blue}// stage one}
			\ELSE
			\STATE $\bm{\beta}^{k,0}_r= \ldots = \bm{\beta}^{k,V}_r \leftarrow$ Initialize based on weight \\ score vector $E^k$  \qquad \qquad \qquad \qquad \emph{\color{blue}// stage two}
			\ENDIF
			\FOR{each iteration $v=0,1,\dots,V-1$}
			\STATE $\theta^{k,v}_r \sim \Tilde{\pi}^{k,v}_r = \bm{\beta}^{k,v}_r \circ \mathcal{N}(U^{k,v}_r, \Tilde{s}^2I)$
			\STATE $U^{k,v+1}_{r} \leftarrow  U^{k,v}_r - \eta \big[\bm{\beta}^{k,v}_r \circ \nabla_U \mathcal{L}^k(\Tilde{\pi}^{k,v}_r)\big]$
			\IF{$r \leq R_b$ \AND $v>\tau$ \AND $v\, \%\, \tau = 0$}
			\STATE Calculate $\Delta\mathcal{L}^{k,v}_r$ using (\ref{loss_diff})
			\IF{$\Delta\mathcal{L}^{k,v}_r > 0 $}
			\STATE $\bm{\beta}^{k,v+1}_r= \ldots = \bm{\beta}^{k,v+\tau}_r \leftarrow$ Randomly \\ sample from $\mathcal{Z}^S_N$
			\ELSE
			\STATE $\bm{\beta}^{k,v+1}_r= \ldots = \bm{\beta}^{k,v+\tau}_r \leftarrow \bm{\beta}^{k,v}_r$
			\ENDIF
			\ENDIF \qquad \qquad \qquad \emph{\color{blue}// only execute in stage one}
			\STATE Update weight score vector $E^k$ by (\ref{score_update})
			\ENDFOR
			\STATE \textbf{return} $\bm{\beta}^{k,V}_r$,  $\{\bm{\mu}_{r,j}^{k,V} | \bm{\mu}_{r,j}^{k,V}\in U_r^{k,V},\,\beta^{k,V}_{r,j} \neq 0\}$
		\end{algorithmic}
		\caption{FedBIAD}
		\label{alg:FedBAD}
	\end{algorithm}
	
	\begin{figure*}
		\begin{center}
			\centerline{\includegraphics[width=1.98\columnwidth]{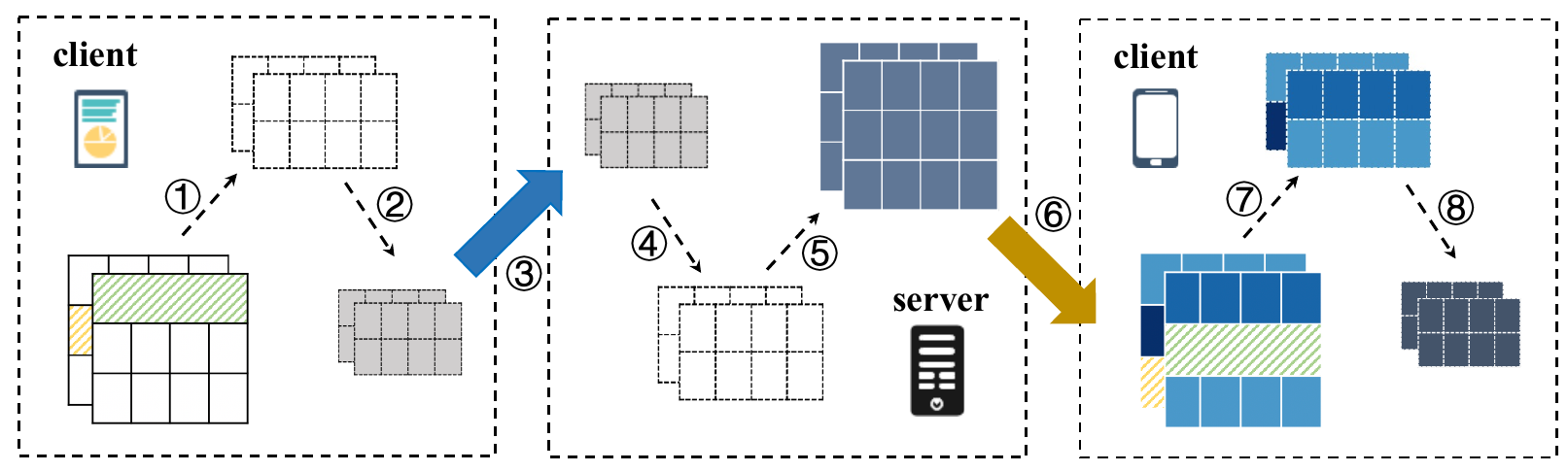}}
			\caption{FedBIAD with a sketched compression - each client (1) drops partial rows, (2) compresses variational parameters of the remaining rows, and (3) uploads compressed variational parameters. The server (4) decompresses and (5) reconstructs weight matrices, and aggregates recovered variational parameters, which are sent to the newly selected clients (6). And then, clients repeat dropping and compression in the next round as (7) and (8). Green and yellow masks indicate that rows are zeroed out.}
			\label{server_client}
		\end{center}
		\vskip -0.2in
	\end{figure*}
	
	\subsection{Experience-based Importance Indicator}
	\label{importance}
	
	We design the weight score vector $E^k=[E_1^k, \ldots, E_\mathrm{J}^k]$ for each client $k$, which is row-wise for fully connected neural networks or RNNs, but filter-wise for CNNs. In the training process described in Section \ref{stage one}, client $k$ records adaptive dropout times of each weight row for fully connected neural networks or RNNs in $E^k$. While for CNN, client $k$ reports the adaptive dropout times of each filter in $E^k$. Assuming that client $k$ holds the $j$-th weight row/filter $\textbf{w}_j$ in the $v$-th iteration, the score $E_j^k$ of $\textbf{w}_j$ is updated iteratively by
	\begin{equation}
		\label{score_update}
		E_j^k = \left\{
		\begin{array}{ll}
			E_j^k + 1, & \text{if}\;\;\Delta\mathcal{L}^{k,v}_r \leq 0, \\
			E_j^k + e_j, & \text{if}\;\;\Delta\mathcal{L}^{k,v}_r > 0.
		\end{array}
		\right.
	\end{equation}
	If the dropping label $\beta^{k,v+1}_j=1$ for the weight row $\textbf{w}_j$, then $e_j=1$; otherwise, $e_j=0$.
	
	As FL progresses until global round $r>R_b$, FedBIAD enters stage two, where the accumulated dropout experiences are enough to be regarded as the importance indicator. This experience-based importance indicator can guide the client to select an effective dropping pattern, which is beneficial for loss reduction. Hence, in stage two, client $k$ can set the dropping pattern based on $E^k$ instead of random sampling, as shown in Fig. \ref{weight_dropout}. The weight score vector determines the dropping pattern as follows. If global round $r>R_b$, client $k$ calculates the threshold $\lambda^{k}_r$ that is the $p$-quantile of $E^k$.  If the score $E^k_j > \lambda^{k}_r$, then the dropping label $\beta^{k,0}_{r,j}=\ldots=\beta^{k,V}_{r,j}=1$ for the $j$-th weight row/filter; otherwise, $\beta^{k,0}_{r,j}=\ldots=\beta^{k,V}_{r,j}=0$. 
	
	\subsection{Global Aggregation}
	\label{aggregation}
	
	After $V$ training iterations on local data, each client $k\in\mathcal{C}_r$ transmits nonzero elements of the latest variational parameters $\{\bm{\mu}_{r,j}^{k,V} | \bm{\mu}_{r,j}^{k,V}\in U_r^{k,V},\,\beta^{k,V}_{r,j} \neq 0,\; \text{for}\; j=1,\dots,\mathrm{J}\}$ and the dropping pattern $\bm{\beta}^{k,V}_r$ to the server. Subsequently, the server reconstructs complete variational parameters $\bm{\beta}^{k,V}_r \circ U_{r}^{k,V}$ based on the received $\bm{\beta}^{k,V}_r$, and performs global aggregation by averaging variational parameters \cite{var_agg_2022}:
	\begin{equation}
		\label{agg}
		U_{r} = \frac{\sum_{k\in\mathcal{C}_r}{|\mathcal{D}^k|\bm{\beta}^{k,V}_r \circ U_{r}^{k,V}}}{\sum_{k\in\mathcal{C}_r}{|\mathcal{D}^k|}}.
	\end{equation}
	Essentially, the global variational approximation $\tilde{\pi}^g_r$ in round $r$ can be denoted as
	\begin{equation}
		\label{global_weights}
		\theta_r \sim \tilde{\pi}^g_r = \mathcal{N}(U_r, \Tilde{s}^2I),
	\end{equation}
	where $\theta_r$ is the global weights in round $r$.
	
	The entire process of FedBIAD is summarized in Algorithm \ref{alg:FedBAD}, and Fig. \ref{server_client} briefly depicts the iterative procedure. Moreover, inspired by \cite{AFD_2021}, we can employ FedBIAD in combination with existing sketched compression methods to further lighten the communication burden. As shown in Fig. \ref{server_client}, variational parameters are compressed in steps (2) and (8).
	
	\subsection{Convergence Analysis}
	\label{convergence_analysis}
	
	Based on Section \ref{variational_bayesian_inference}, FedBIAD explores a global variational approximation $\Tilde{\pi}^g$ to estimate the tempered posteriors of client-side data. Considering a global model $f_\theta$ with $L$ layers whose weights $\theta$ follow $\Tilde{\pi}^g$, we define the unsparse number $S$ as the number of nonzero weights in $\theta$ and $D$ as the dimension of hidden layers in the global model. In this way, the global model is determined by $(S,L,D)$, and $\Theta_{S,L,D}$ represents the solution space composed of all feasible weights. The input dimension of the global model is $d$. As discussed in Section \ref{problem_formulatiom}, each device $k\in\mathcal{K}$ owns its data $\mathcal{D}^k$ and unknown true function $f_0^k$ such that $y^k=f^k_0(\bm{x}^k)$ for any $(\bm{x}^k, y^k)\in \mathcal{D}^k$.
	
	We define $m$ as the client-side total input data, which is associated with the training round $r$. Considering that different clients have different amounts of local data, after $r$ rounds,  the minimum amount of client-side total input data is denoted as $m_r=r\times V \times \min\{|\mathcal{D}^1|, \ldots, |\mathcal{D}^K|\}$, where $V$ is the number of training iterations in each round. To analyze the convergence of the average generalization error in FedBIAD, we start with some assumptions.
	
	\begin{assumption}
		The activation functions $\rho$, $\varrho$, and $\phi$ are $1$-Lipschitz continuous.
		\label{ass:ass1}
	\end{assumption}
	\begin{assumption}
		The absolute values of all weights in optimal global model $\theta^*$ have an upper bound of $B \geq 2$.
		\label{ass:ass2}
	\end{assumption} 
	Assumption \ref{ass:ass1} and \ref{ass:ass2} are common in Bayesian convergence analysis works \cite{Convergence_Bayes_2020, pFedBayes_2022}, which are realistic. The activation functions (i.e., relu function $\rho$, tanh function $\varrho$, sigmoid function $\phi$) we used are actually 1-Lipschitz continuous. Besides, the optimal global model is denoted by $ \theta^{\ast}=\mathrm{arg} \min_{\theta \in \Theta_{S,L,D}} \frac{1}{K} \sum_{k=1}^K\|f_{\theta}-f^k_0\|_2^2$, which is fixed. Hence, the absolute values of all weights in $\theta^*$ are fixed, and they must have an upper bound.
	
	Based on \cite{Convergence_Bayes_2020, pFedBayes_2022} and Definition \ref{def:def1}, the average generalization error of global model learned by FedBIAD is the expected average of the squared L2-distance between the global model $f_\theta$ and local unknown true functions $\{f_0^k|k\in \mathcal{K}\}$, denoted by:
	\begin{equation}
		\label{avg_error}
		\mathbb{E} \bigg[ \frac{1}{K} \sum_{k=1}^{K} \int \|{f_{\theta}-f^k_0}\|^2_2 \Tilde{\pi}^g(\mathrm{d}\theta) \bigg].
	\end{equation}
	The constant posterior variance is set to
	\begin{align}
		\label{same_variance}
		\Tilde{s}^2 & = \frac{S}{16m d^2} \log(3D)^{-1} (2BD)^{-2L} \\ 
		& \bigg\{ \Big(d+1+\frac{1}{BD-1} \Big)^2 + \frac{1}{(BD)^2-1} + \frac{2}{(BD-1)^2} \bigg\}^{-1}.\notag
	\end{align}
	In FedBIAD, the posterior variances of local variational approximations on clients and the global variational approximation on the server are the same in each round. Hence, clients and the server do not exchange posterior variances $\Tilde{s}^2$ and only compute $\Tilde{s}^2$ via (\ref{same_variance}), which is beneficial for communication efficiency. Based on the setting of posterior variance in (\ref{same_variance}), we can derive the upper bound of the average generalization error expressed by (\ref{avg_error}) for FedBIAD in Theorem \ref{thm:upper_bound}.
	\begin{theorem}
		\label{thm:upper_bound}
		Let Assumption \ref{ass:ass1} and \ref{ass:ass2} hold, considering different numbers of local data in different clients such that the minimum amount of client-side total input data up to round $r$ is denoted as $m_r=r\times V \times \min\{|\mathcal{D}^1|, \ldots, |\mathcal{D}^K|\}$, for any $\alpha \in (0,1)$, we obtain the average generalization error of global model $f_{\theta}$ learned by FedBIAD in the server:
		\begin{align}
			\label{upper_bound}
			\mathbb{E} \bigg[ \frac{1}{K} & \sum_{k=1}^{K} \int \|{f_{\theta}-f^k_0}\|^2_2 \Tilde{\pi}^g(\mathrm{d}\theta) \bigg] \notag\\
			& \leq  \frac{2\sigma^2}{\alpha(1-\alpha)}(1+ \frac{\alpha}{\sigma^2})\varepsilon_{m_r}^{S,L,D} + \frac{2}{K(1-\alpha)} \sum_{k=1}^{K} \xi^k
		\end{align}
		with
		\begin{align}
			\label{varepsilon_m}
			\varepsilon_{m_r}^{S,L,D} = &  \frac{SL}{m_r}\log(2BD) + \frac{3S}{m_r}\log(LD) + \frac{SB^2}{2m_r} \notag \\ 
			& + \frac{2S}{m_r} \log\Big(4d\max(\frac{m_r}{S},1)\Big)
		\end{align}
		and
		\begin{equation}
			\label{xi_k}
			\xi^k = \inf_{\theta^{\ast}\in \Theta_{S,L,D}}\|f_{\theta^{\ast}} - f^k_0\|_\infty^2,
		\end{equation}where $\theta$ is the global weights computed by (\ref{global_weights}) and $\sigma^2$ is the likelihood variance defined in Section \ref{variational_bayesian_inference}.
	\end{theorem}
	
	Through Theorem \ref{thm:upper_bound}, we get the upper bound of the average generalization error of global model, which can ensure convergence of FedBIAD. Referring to \cite{Convergence_Bayes_2020}, if local true functions $\{f^k_0|k\in\mathcal{K}\}$ are actually neural networks with structure $(S, L, D)$, the term $\xi^k$ denoted by (\ref{xi_k}) on the right-hand side of (\ref{upper_bound}) vanishes. Then, we only need to analyze the first term on the right-hand side of (\ref{upper_bound}), where $\varepsilon_{m_r}^{S,L,D}$ is calculated by (\ref{varepsilon_m}). According to (\ref{varepsilon_m}),  $\varepsilon_{m_r}^{S,L,D}$ apparently declines with the increasing of client-side total input data $m_r$. While $m_r$ increases with the growth of training round $r$. Therefore, as the round grows, the upper bound of the generalization error of FedBIAD decreases, and FedBIAD gradually converges.
	
	However, if local true functions $\{f^k_0|k\in\mathcal{K}\}$ are not neural networks with structure $(S, L, D)$ like the global model $f_\theta$, we have to consider the term $\xi^k$. Motivated by \cite{pFedBayes_2022, lemma5.1_2018}, we assume that $\{f^k_0|k\in\mathcal{K}\}$ are $\gamma$-H\"older smooth functions with $0<\gamma<d$. According to Lemma 5.1 in \cite{lemma5.1_2018} and Corollary 3 in \cite{Convergence_Bayes_2020}, there exist constants $C,\,C'$ such that $\xi^k \leq C m^{\frac{-2\gamma}{2\gamma+d}}_r\cdot\log^2m_r$ and $\varepsilon_{m_r}^{S,L,D} \leq C' m^{\frac{-2\gamma}{2\gamma+d}}_r\cdot\log^2m_r$. Then, we have
	\begin{equation}
		\label{xiaoyu}
		\mathbb{E} \bigg[\frac{1}{K} \sum_{k=1}^{K} \int \|{f_{\theta}-f^k_0}\|^2_2 \Tilde{\pi}^g(\mathrm{d}\theta) \bigg] \leq C_1 m_r^{\frac{-2\gamma}{2\gamma+d}}\cdot \log^2m_r,
	\end{equation}where $C_1 > 0$ is a constant. In addition, based on Theorem 8 in \cite{ada_generalization_2020}, there exists a constant $C_2 > 0$ such that 
	\begin{equation}
		\label{dayu}
		\inf_\theta \sup_{\{f^k_0\}_{k=1}^K} \frac{1}{K}\sum_{k=1}^{K}\int \|{f_{\theta}-f^k_0}\|^2_2 \Tilde{\pi}^g(\mathrm{d}\theta) \geq C_2 m_r^{\frac{-2\gamma}{2\gamma+d}}.
	\end{equation}The (\ref{xiaoyu}) gives an upper bound of generalization error for FedBIAD with $\gamma$-H\"older smooth functions, while (\ref{dayu}) provides the minimax lower bound \cite{ada_generalization_2020, pFedBayes_2022} of generalization error for FedBIAD. Both (\ref{xiaoyu}) and (\ref{dayu}) have the same term of $m_r^{\frac{-2\gamma}{2\gamma+d}}$, through which we can derive the average generalization error of FedBIAD converges at the \emph{minimax rate $m_r^{\frac{-2\gamma}{2\gamma+d}}$ up to a squared logarithmic factor} for the expected $L_2$-distance.
	
	Note that, the setting of posterior variance can affect the generalization error bound. The optimal variance is derived in (\ref{same_variance}), which is used to derive the minimum generalization error bound in (\ref{upper_bound}). Furthermore, FedBIAD only exchanges variational model parameters without sharing client-side data. Therefore, it does not introduce extra privacy concerns and existing FL privacy-preserving methods can be directly applied.
	
	\section{Experiment Evaluation}
	
	In this section, extensive experiments are conducted on five datasets for image classification and next-word prediction tasks to evaluate the performance of FedBIAD compared to the state-of-the-art methods.
	
	\subsection{Experimental Settings}
	
	\textbf{Datasets.} We consider five datasets for the experiment evaluation: MNIST \cite{Mnist_dataset}, Fashion-MNIST (FMNIST) \cite{FMNIST_dataset}, Penn TreeBank (PTB) \citep{ptb_1993}, WikiText-2 \citep{wiki2_2016}, and Reddit \citep{leaf_benchmark_2018}. 
	
	MNIST and FMNIST are 10-class image datasets containing
	60,000 training data and 10,000 test data. MNIST is for digital handwriting classification, while FMNIST consists of pictures of fashion clothes, and its task is more challenging than simple MNIST. For MNIST and FMNIST, the number of clients is 1000, and we utilize the non-IID partitioning strategy in \cite{pFedBayes_2022}. 
	
	PTB, WikiText-2, and Reddit are three English datasets widely used for next-word prediction \cite{leaf_benchmark_2018, fedatt_2019}. The PTB has a 5.1M training corpus, smaller than most modern datasets. The WikiText-2 is over 2 times larger than the PTB, featuring a vocabulary of more than 30,000 words. Both of them are IID. We randomly sample data without overlap and allocate them to 100 clients. For Reddit, the data are non-IID and the top 100 users with more data are chosen as clients, so that different clients have different sample sizes. The data in clients are split into the training set, validation set, and test set.
	
	\noindent \textbf{Baselines.} We compare with the following baselines:
	\begin{itemize}
		\item[$\bullet$] FedAvg \cite{fedavg_2017}: FedAvg is a pioneering work of FL, where each participating client periodically transmits all weights of the local model to the server.
		\item[$\bullet$] FedDrop \cite{fed_dropout_2019}: Each client randomly drops partial units of DNN, which applies to convolutional and Fully Connected (FC) layers and does not extend to recurrent layers. 
		\item[$\bullet$] AFD \cite{AFD_2021}: AFD builds upon FedDrop and proposes to train and transmit only the necessary weights that are not affected by the dropout, but the dropout is only applied to non-recurrent connections of models.
		\item[$\bullet$] FedMP \cite{FedMP_2022}:  FedMP assumes that small weights have a weak effect on model accuracy, where each client prunes weights with lower absolute values.
		\item[$\bullet$] Fjord \cite{fjord_2021}: Each client extracts the lower footprint submodel by ordered dropout, which preferentially drops the right-most adjacent neurons of each layer. The left-most neurons are used by more clients during training.
		\item[$\bullet$] HeteroFL \cite{heterofl_2021}: HeteroFL reduces the number of local weights by shrinking the width of hidden layers, where different clients could adopt different shrinkage ratios.
	\end{itemize}
	
	\noindent \textbf{Simulation Parameters.} In terms of parameter settings, we mainly refer to \cite{param_settings_1_2020} for image classification tasks and \cite{regularizing_2017, fedatt_2019} for next-word prediction tasks. Concretely, we adopt a fully connected model that uses a hidden layer, a ReLU activation function, and a softmax layer for image classification. The number of hidden units is set to 128 on MNIST and 256 on FMNIST. We conduct various experiments with stochastic gradient descent (SGD) optimizer. Under next-word prediction tasks, the model consists of an embedding layer with 300 units, a two-layer LSTM with 300 hidden units, and a Fully Connected (FC) layer. We use the SGD optimizer with the clipped gradient norm. The iteration interval is set to $\tau=3$, and the client selection ratio is $\kappa=0.1$. All methods execute 60 global rounds, and the boundary of the stage is $R_b=55$.
	
	\begin{table}
		\centering
		\caption{Test accuracy (i.e., `Acc') and per-round upload parameter size (i.e., `Upload Size') of different methods.}
		\setlength{\tabcolsep}{2.3mm}{
			\begin{tabular}{c|c|c|c|c}
				\toprule[1pt]
				\textbf{Dataset}&\textbf{Method}&\textbf{Acc (\%)}&\textbf{Upload Size}&\textbf{Save Ratio} \\ \midrule[1pt]
				\multirow{7}*{MNIST}& FedAvg & 95.06\scalebox{0.8}{$\pm$0.03} & 531KB & 1 $\times$ \\
				& FedDrop & 95.03\scalebox{0.8}{$\pm$0.05} & 424KB & 1.25 $\times$ \\
				& AFD & 94.49\scalebox{0.8}{$\pm$0.10} & 424KB & 1.25 $\times$ \\
				& FedMP &  95.09\scalebox{0.8}{$\pm$0.03} & 477KB & 1.10 $\times$ \\
				& FjORD & 94.93\scalebox{0.8}{$\pm$0.08} & 437KB &  1.21 $\times$ \\
				& HeteroFL &  94.98\scalebox{0.8}{$\pm$0.06} & 432KB & 1.23 $\times$ \\
				& \textbf{FedBIAD} & \textbf{95.20}\scalebox{0.8}{$\pm$0.11} & \textbf{424KB} & \textbf{1.25 $\times$} \\
				\midrule[0.5pt]
				\multirow{7}*{FMNIST}& FedAvg & 81.18\scalebox{0.8}{$\pm$1.09} & 1.1MB & 1 $\times$ \\
				& FedDrop & 81.12\scalebox{0.8}{$\pm$0.47} & 530KB & 2 $\times$ \\
				& AFD & 82.37\scalebox{0.8}{$\pm$0.37} & 530KB & 2 $\times$ \\
				& FedMP & 82.40\scalebox{0.8}{$\pm$0.26} & 862KB & 1.3 $\times$ \\
				& FjORD & 82.64\scalebox{0.8}{$\pm$0.10} & 718KB &  1.5 $\times$ \\
				& HeteroFL & 82.68\scalebox{0.8}{$\pm$0.51} & 685KB & 1.6 $\times$ \\
				& \textbf{FedBIAD} & \textbf{83.59}\scalebox{0.8}{$\pm$0.13} & \textbf{530KB} & \textbf{2 $\times$} \\
				\midrule[0.5pt]
				\multirow{7}*{PTB}& FedAvg &  28.54\scalebox{0.8}{$\pm$0.15}  & 29.8MB & 1 $\times$ \\
				& FedDrop & 27.81\scalebox{0.8}{$\pm$0.21} & 23.8MB & 1.25 $\times$ \\
				& AFD & 28.67\scalebox{0.8}{$\pm$0.03} & 22.4MB & 1.3 $\times$ \\
				& FedMP & 28.76\scalebox{0.8}{$\pm$0.20} & 22.7MB & 1.3 $\times$ \\
				& FjORD & 27.88\scalebox{0.8}{$\pm$0.18} & 21.4MB &  1.4 $\times$ \\
				& HeteroFL & 26.80\scalebox{0.8}{$\pm$0.42} & 20.4MB & 1.5 $\times$ \\
				& \textbf{FedBIAD} & \textbf{29.85}\scalebox{0.8}{$\pm$0.15} & \textbf{16.4MB} & \textbf{2 $\times$} \\
				\midrule[0.5pt]
				\multirow{7}*{WikiText-2}& FedAvg & 31.86\scalebox{0.8}{$\pm$0.05}  & 75.3MB & 1 $\times$ \\
				& FedDrop & 32.02\scalebox{0.8}{$\pm$0.55} & 57.9MB & 1.3 $\times$ \\
				& AFD & 31.20\scalebox{0.8}{$\pm$0.64} & 56.5MB & 1.3 $\times$ \\
				& FedMP & 32.53\scalebox{0.8}{$\pm$1.48} & 59.1MB &  1.3 $\times$ \\
				& FjORD & 31.16\scalebox{0.8}{$\pm$0.03} & 54.0MB & 1.4 $\times$ \\
				& HeteroFL & 31.84\scalebox{0.8}{$\pm$0.64} & 52.9MB & 1.4 $\times$ \\
				& \textbf{FedBIAD} & \textbf{33.16}\scalebox{0.8}{$\pm$0.97} & \textbf{39.1MB} & \textbf{2 $\times$} \\
				\midrule[0.5pt]
				\multirow{7}*{Reddit}& FedAvg & 31.68\scalebox{0.8}{$\pm$0.54} & 29.8MB & 1 $\times$ \\
				& FedDrop & 31.84\scalebox{0.8}{$\pm$0.08} & 24.1MB & 1.25 $\times$ \\
				& AFD & 32.26\scalebox{0.8}{$\pm$0.11} & 22.5MB & 1.3 $\times$ \\
				& FedMP & 31.06\scalebox{0.8}{$\pm$0.19} & 22.7MB &  1.3 $\times$ \\
				& FjORD & 31.35\scalebox{0.8}{$\pm$0.38} & 21.4MB &  1.4 $\times$ \\
				& HeteroFL & 31.24\scalebox{0.8}{$\pm$0.11} & 20.4MB & 1.5 $\times$ \\
				& \textbf{FedBIAD} & \textbf{33.93}\scalebox{0.8}{$\pm$0.28} & \textbf{16.4MB} & \textbf{2 $\times$} \\
				\bottomrule[1pt]
		\end{tabular}}
		\label{tab:main_results}
		\vskip -0.05in
	\end{table}
	
	\textbf{Dropout rate settings.} Referring to \cite{AFD_2021}, empirical dropout rates are between 0.1 to 0.5, which are associated with model scales. Different datasets require different-scale models \cite{federated_dropout_2021}. For example, PTB is more challenging than simple MNIST, so the model for training PTB is bigger than MNIST. Large-scale models are often possible to use higher dropout rates \cite{AFD_2021}. For small-scale models (such as 531KB for MNIST), high rates make shrinking models unable to extract accurate features, degrading accuracy \cite{federated_dropout_2021}. For MNIST, if dropout rate $p=0.5$, the accuracy results of FedDrop, AFD, FedMP, and \emph{FedBIAD} are 94.45, 94.26, 94.45, and \emph{94.69}, lower than $p=0.2$ (see Table \ref{tab:main_results}). Thus, different dropout rates are necessary for different datasets requiring different model scales. Motivated by \cite{federated_dropout_2021,AFD_2021}, we set lower dropout rate $p=0.2$ for MNIST with small-scale model ($<1$M), while other four datasets requiring large-scale models ($\geq 1$M) use the same dropout rate $p=0.5$. This selection strategy for dropout rates is feasible in practice.
	
	\subsection{Experimental Results}
	\label{experimental_results}
	
	\begin{table*}[tbp]
		\caption{The comparison between sketched compression methods and FedBIAD combined with DGC on five datasets.}
		\label{tab:DGC_Results}
		\centering
		\begin{threeparttable}
			\renewcommand{\multirowsetup}{\centering}
			\setlength{\tabcolsep}{5pt}
			\begin{tabular}{c|c|ccccccc}
				\toprule[1pt]
				\multicolumn{2}{c}{Methods / Metrics}  & FedPAQ \cite{FedPAQ_2020} & SignSGD \cite{signSGD_2018} & STC \cite{STC_2020} & DGC \cite{dgc_2018} & AFD+DGC \cite{AFD_2021} & Fjord+DGC \cite{fjord_2021} & \textbf{FedBIAD+DGC}  \\ 
				\midrule[1pt]
				\multirow{3}{*}{{MNIST}}  & Accuracy (\%) & 94.90 \scalebox{0.8}{$\pm$0.09} & 92.04 \scalebox{0.8}{$\pm$0.55} & 90.56\scalebox{0.8}{$\pm$0.17} & 94.84\scalebox{0.8}{$\pm$0.11} & 94.39\scalebox{0.8}{$\pm$0.20} & 94.93\scalebox{0.8}{$\pm$0.03} & \textbf{95.22}\scalebox{0.8}{$\pm$0.11}\\
				& Upload size & 129KB & 16KB & 3KB & 3KB & 2KB & 2KB & 2KB \\
				& Save ratio & 4 $\times$ & 33 $\times$ & 177 $\times$ & 177 $\times$ & 265 $\times$ & 265 $\times$ & 265 $\times$\\
				\midrule[0.5pt]
				\multirow{3}{*}{{FMNIST}}  & Accuracy (\%) & 78.64\scalebox{0.8}{$\pm$0.02} & 76.57\scalebox{0.8}{$\pm$0.43} & 81.13\scalebox{0.8}{$\pm$0.43} & 80.64\scalebox{0.8}{$\pm$0.09} & 81.96\scalebox{0.8}{$\pm$0.43} & 82.16\scalebox{0.8}{$\pm$0.56} & \textbf{82.96}\scalebox{0.8}{$\pm$0.03}\\
				& Upload Size & 258KB & 33KB & 6KB & 4KB & 3KB & 3KB & 3KB \\
				& Save Ratio & 4 $\times$ & 34 $\times$ & 188 $\times$ & 281 $\times$ & 375 $\times$ & 375 $\times$ & 375 $\times$\\
				\midrule[0.5pt]
				\multirow{3}{*}{PTB}  & Accuracy (\%) & 28.60\scalebox{0.8}{$\pm$0.01}  & 23.76\scalebox{0.8}{$\pm$0.24} & 24.42\scalebox{0.8}{$\pm$0.46} & 28.10\scalebox{0.8}{$\pm$0.33} & 27.74\scalebox{0.8}{$\pm$0.10}  & 27.50\scalebox{0.8}{$\pm$0.24} & \textbf{28.77}\scalebox{0.8}{$\pm$0.05}\\
				& Upload Size & 7.1MB & 908KB & 148KB & 95KB & 71KB & 71KB & \textbf{53KB} \\
				& Save Ratio & 4 $\times$ & 33 $\times$ & 206 $\times$ & 321 $\times$ & 429 $\times$ & 429 $\times$ & \textbf{575} $\times$\\
				\midrule[0.5pt]
				\multirow{3}{*}{{WikiText-2}}  & Accuracy (\%) & 32.04\scalebox{0.8}{$\pm$0.02} & 30.62\scalebox{0.8}{$\pm$0.42} & 28.92 \scalebox{0.8}{$\pm$0.91} & 31.58\scalebox{0.8}{$\pm$0.06} & 31.24\scalebox{0.8}{$\pm$0.16} & 30.92\scalebox{0.8}{$\pm$0.25} & \textbf{33.78}\scalebox{0.8}{$\pm$0.43}\\
				& Upload Size & 18.8MB & 2.4MB & 374KB & 215KB & 180KB & 179KB & \textbf{126KB} \\
				& Save Ratio & 4 $\times$ & 32 $\times$ & 206 $\times$ & 359 $\times$ & 428 $\times$ & 430 $\times$ & \textbf{612} $\times$\\
				\midrule[0.5pt]
				\multirow{3}{*}{{Reddit}}  & Accuracy (\%) & 32.36\scalebox{0.8}{$\pm$0.13} & 29.86\scalebox{0.8}{$\pm$0.06} & 30.22\scalebox{0.8}{$\pm$0.08} & 31.23\scalebox{0.8}{$\pm$0.48} & 32.19\scalebox{0.8}{$\pm$0.28} & 30.85\scalebox{0.8}{$\pm$0.28} & \textbf{32.51}\scalebox{0.8}{$\pm$0.06}\\
				& Upload Size & 7.1MB & 960KB & 148KB & 97KB & 88KB &  86KB & \textbf{52KB} \\
				& Save Ratio & 4 $\times$ & 32$\times$ & 206 $\times$ & 314 $\times$ & 346 $\times$ &  355 $\times$ & \textbf{587} $\times$\\
				\bottomrule[1pt]
			\end{tabular}
		\end{threeparttable}
		\vskip -0.1in
	\end{table*}
	
	\begin{figure*}[htbp]
		\centering
		\subfloat[Training loss and test accuracy vary with rounds on MNIST.]{
			\begin{minipage}[t]{0.49\linewidth}
				\centering
				\centerline{\includegraphics[width=0.95\columnwidth]{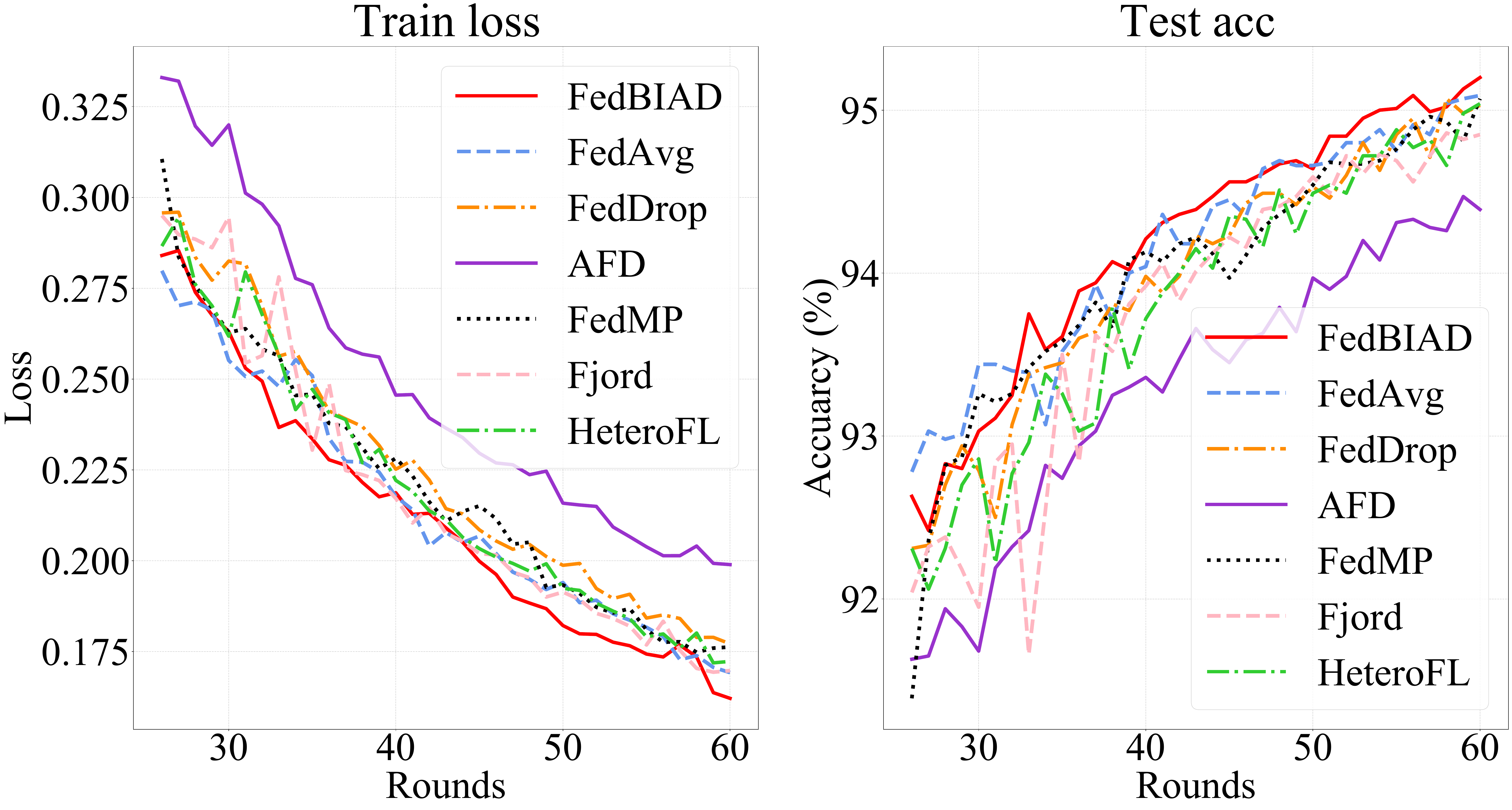}}
				\label{convergence_mnist}
			\end{minipage}%
		}%
		\subfloat[Training loss and test accuracy vary with rounds on WikiText-2.]{
			\begin{minipage}[t]{0.49\linewidth}
				\centering
				\centerline{\includegraphics[width=0.92\columnwidth]{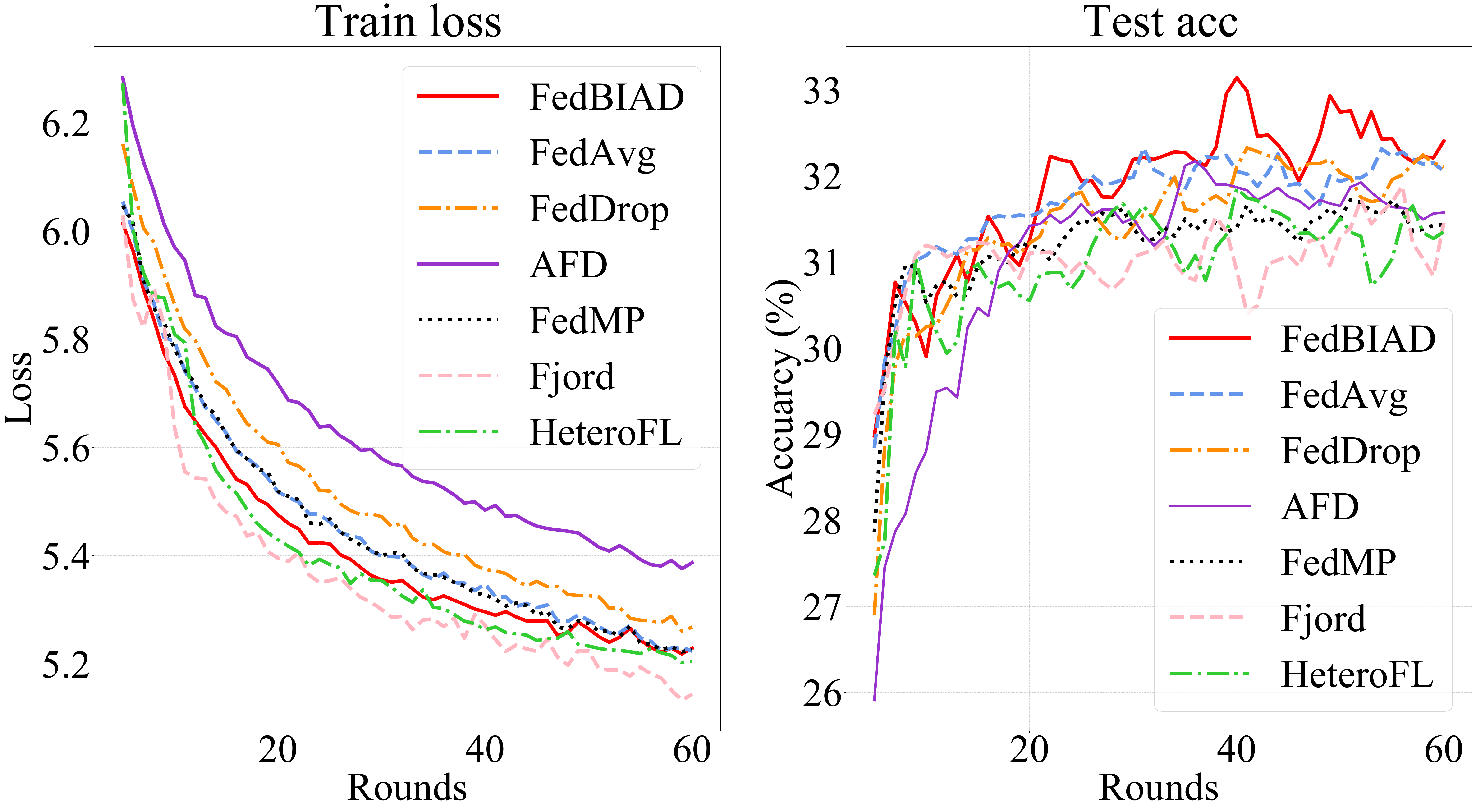}}
				\label{convergence_wiki}
			\end{minipage}
		}
		\centering
		\caption{Training loss and test accuracy versus rounds on MNIST and WikiText-2.}
		\label{convergence_curves}
		\vskip -0.1in
	\end{figure*}
	
	\textbf{Performance Comparison.} Table \ref{tab:main_results} shows the performance results of different methods on test data, where `Save Ratio' is the saving multiples in upload parameter size per round compared to FedAvg, and the results are obtained in a NVIDIA Tesla V100 GPU with 32GB. Besides, `Upload Size' in Table \ref{tab:main_results} has included the size of transmitted binary dropping pattern $\bm{\beta}$. Although each hidden layer has an independent row/filter size, the number of dropping labels for these rows/filters in all hidden layers is much less than the number of model weights. Besides, each dropping label is 1bit, while each weight is 32bit. Therefore, the size of $\bm{\beta}$ is much smaller than the model weights. For example, $\bm{\beta}$ in the Reddit dataset is 0.3KB, much smaller than the original model size of 29.8MB. Note that, for next-word prediction, we follow \cite{FL_keyboard_pre_2018} to regard top-3 accuracy as the evaluation metric of model performance because mobile keyboards generally include three candidates, while the evaluation metric of the image classification is top-1 accuracy. 
	As shown in Table \ref{tab:main_results}, FedBIAD achieves consistent state-of-the-art accuracy with minimal per-round upload parameters across all datasets. For image classification, FedBIAD reduces 2$\times$ uplink costs while improving accuracy by 2.41\% on FMNIST compared to FedAvg. In terms of MNIST and FMNIST, the model architecture consists of several FC layers. Because both FedDrop and AFD can apply dropout to all FC layers, FedBIAD does not significantly outperform FedDrop and AFD in communication efficiency. However, for MNIST, FedBIAD increases accuracy by 0.17\% and 0.71\% compared to FedDrop and AFD. Similarly, FedBIAD shows 2.47\% and 1.22\% accuracy improvements on FMNIST.
	
	We also observe from Table \ref{tab:main_results} that on next-word prediction, FedBIAD outperforms other baselines in both uplink communication efficiency and test accuracy, achieving a 2$\times$ reduction of uplink costs and up to 1.45\% accuracy increase compared to FedAvg. Because FedDrop and AFD cannot be applied to recurrent connections of RNN models, both show smaller save ratios, meaning less reduction in uplink overhead. FedBIAD adaptively drops partial weight rows for recurrent connections, which provides more uplink savings with higher accuracy. Thus, FedBIAD has better communication efficiency than AFD and FedDrop when using LSTM (a typical RNN) layers on next-word prediction. Compared with FedDrop, FedBIAD provides a 1.6$\times$ reduction in uplink costs while increasing accuracy by 1.37\% on Reddit. In comparison to AFD, FedBIAD compresses uplink costs by 1.5$\times$ with 0.97\% accuracy improvement. Besides, FedBIAD reduces 1.4$\times$ uplink costs and is 2.17\% more accurate than Fjord on Reddit.
	
	\begin{figure*}
		\centering
		\subfloat[LTTR on image classification.]{
			\begin{minipage}[t]{0.24\linewidth}
				\centering
				\centerline{\includegraphics[width=0.99\columnwidth]{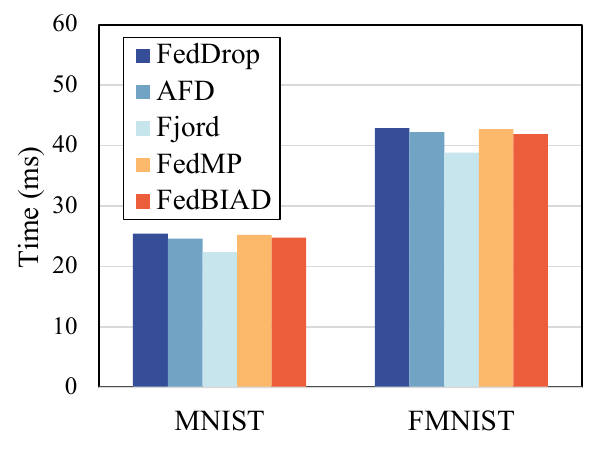}}
				\label{runtime_image}
			\end{minipage}%
		}%
		\subfloat[LTTR on next-word prediction.]{
			\begin{minipage}[t]{0.24\linewidth}
				\centering
				\centerline{\includegraphics[width=0.99\columnwidth]{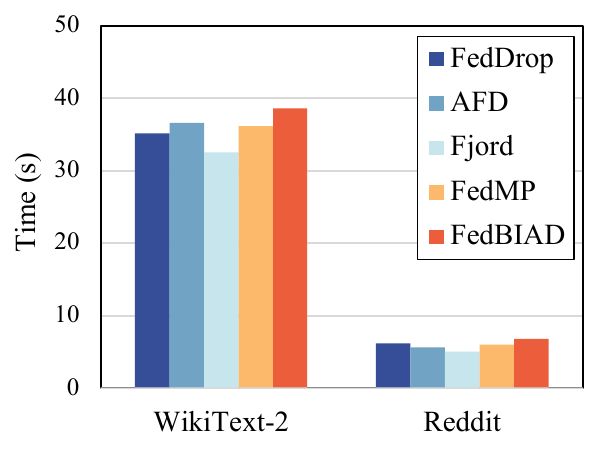}}
				\label{runtime_word}
			\end{minipage}
		}
		\subfloat[TTA on image classification.]{
			\begin{minipage}[t]{0.24\linewidth}
				\centering
				\centerline{\includegraphics[width=0.99\columnwidth]{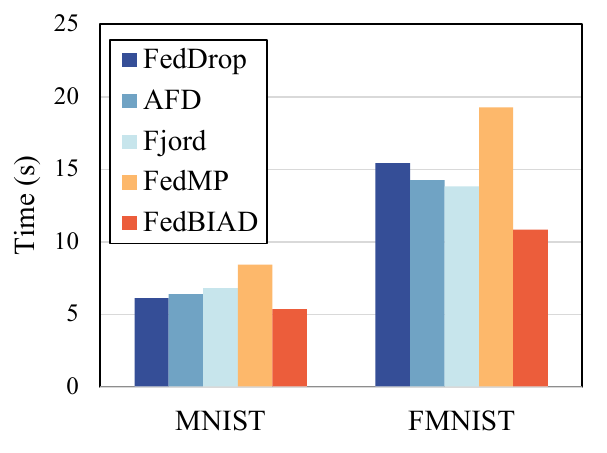}}
				\label{com_time_image}
			\end{minipage}%
		}%
		\subfloat[TTA on next-word prediction.]{
			\begin{minipage}[t]{0.24\linewidth}
				\centering
				\centerline{\includegraphics[width=0.99\columnwidth]{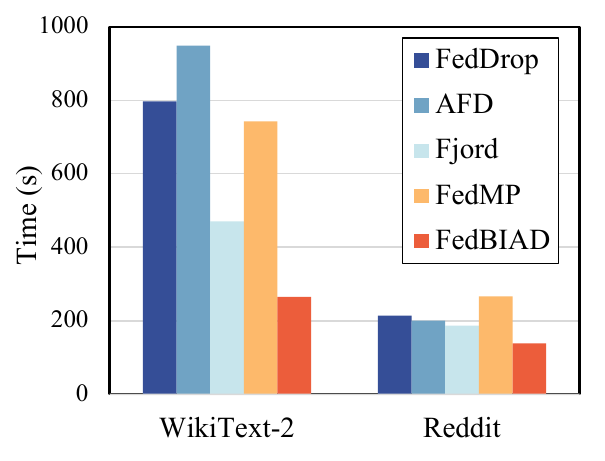}}
				\label{con_time_word}
			\end{minipage}
		}
		\centering
		\caption{Comparison of Local Training Time in a Round (LTTR) and Time-To-Accuracy (TTA) on different datasets.}
		\label{time_analysis}
		\vskip -0.1in
	\end{figure*}
	
	Motivated by \cite{AFD_2021}, we empirically combine FedBIAD with a sketched compression method, DGC \cite{dgc_2018}, further alleviating communication overhead. We also compare FedBIAD with DGC against other sketched compression approaches, among which \emph{SignSGD} \cite{signSGD_2018} and \emph{FedPAQ} \cite{FedPAQ_2020} compress parameters via 1-bit and 8-bit quantizers, respectively. \emph{STC} \cite{STC_2020} integrates sparsification and quantization into a compression framework. Both STC and DGC require clients to upload the positions of compressed parameters. For fairness, we adopt the same position representation approach, where the position representation of each parameter occupies 64 bits \cite{dgc_2018}. The performance results are presented in Table \ref{tab:DGC_Results}. We notice that FedBIAD with DGC consistently achieves the highest accuracy with the least uplink overhead than other methods. Compared to FedAvg, which does not involve any compression strategies, FedBIAD with DGC achieves up to 612$\times$ reduction in uplink overhead while improving accuracy by up to 2.43\%. In addition, FedBIAD with DGC gives up to 3$\times$ better uplink compression than STC with up to 8.07\% accuracy improvement. The uplink communication volume of FedBIAD with DGC is about 2$\times$ less than that of naive DGC, and FedBIAD with DGC is 2.26\% more accurate than naive DGC. Furthermore, we combine AFD, Fjord with DGC, respectively. Both show lower model accuracy and more uplink costs than FedBIAD with DGC.

	\textbf{Convergence Comparison.}  To evaluate the convergence of FedBIAD, we report the average training loss and test accuracy of MNIST and WikiText-2 datasets varying with rounds in Fig. \ref{convergence_curves}. Fig. \ref{convergence_mnist} shows the convergence curves of MNIST. We observe that FedBIAD quickly converges to higher accuracy than other baselines. Additionally, FedBIAD nearly achieves the lowest training loss and the highest test accuracy in each round on MNIST. Fig. \ref{convergence_wiki} presents the convergence curves of WikiText-2. For clarity, the curves are smoothed by the moving average, which will cause a few precision biases. As shown in Fig. \ref{convergence_wiki}, although FedBIAD does not provide the lowest training loss, it converges to the best test accuracy, which indicates that FedBIAD avoids overfitting on training data and outperforms other baselines on test data.
	
	\subsection{Training Time Analysis}
	
	The Local Training Time in a Round (LTTR) can be used to characterize the local computation costs. Fig. \ref{runtime_image} and Fig. \ref{runtime_word} show the LTTR of different datasets, in which time is gained from the end-user device of Macbook Pro 2019 with 32G RAM. We note that FedBIAD increases local computation costs, which is apparent on WikiText-2. Concretely, FedBIAD increases LTTR by 3.4 seconds (8.81\%), 1.99 seconds (5.16\%), and 6.1 seconds (15.81\%) on WikiText-2 compared to FedDrop, AFD, and Fjord.
	
	FedBIAD reduces uplink communication costs with accuracy improvement, so it can accelerate global models to achieve target accuracy. As shown in Fig. \ref{convergence_mnist} and Fig. \ref{convergence_wiki}, under the same number of rounds, FedBIAD can generally achieve higher accuracy, which means that FedBIAD consumes much fewer communication rounds to reach a pre-defined target accuracy. For example, FedBIAD can obtain an accuracy of 31\% after 4 rounds on WikiText-2, while Fjord needs 7 rounds and AFD needs 13 rounds. Motivated by \cite{HACCS_2022}, we utilize Time-To-Accuracy (TTA) to measure the total training time required to achieve target accuracy, which comprises local running time, parameter transmission time, and parameter aggregation time. 
	
	To simulate parameter transmission time, we utilize the T-Mobile 5G network with a download speed of 110.6 Megabits per second (Mbps) and an upload speed of 14.0 Mbps \cite{down_up}. The target accuracies are 90\%, 80\%, 31\%, and 30\% for MNIST, FMNIST, WikiText-2, and Reddit, respectively. Then, we calculate the TTA of different datasets, and the results are shown in Fig. \ref{com_time_image} and Fig. \ref{con_time_word}. We observe that on four datasets, FedBIAD consistently takes the shortest time to achieve the target accuracy compared with baselines. Although FedBIAD increases LTTR by 6.1 seconds on WikiText-2, it reduces uplink costs and the number of rounds, so the TTA of FedBIAD is 205.1 seconds shorter than Fjord. Specifically, FedBIAD reduces 66.67\%, 72.00\%, and 43.61\% TTA on WikiText-2 compared to FedDrop, AFD, and Fjord, respectively. For FMNIST, the reductions are 29.64\%, 23.91\%, and 21.54\%.
	
	\subsection{Effect of Dropout rate}
	
	\begin{figure}
		\centering
		\vskip -0.1in
		\subfloat[Test accuracy versus dropout rate.]{
			\begin{minipage}[t]{0.5\linewidth}
				\centering
				\centerline{\includegraphics[width=0.9\columnwidth]{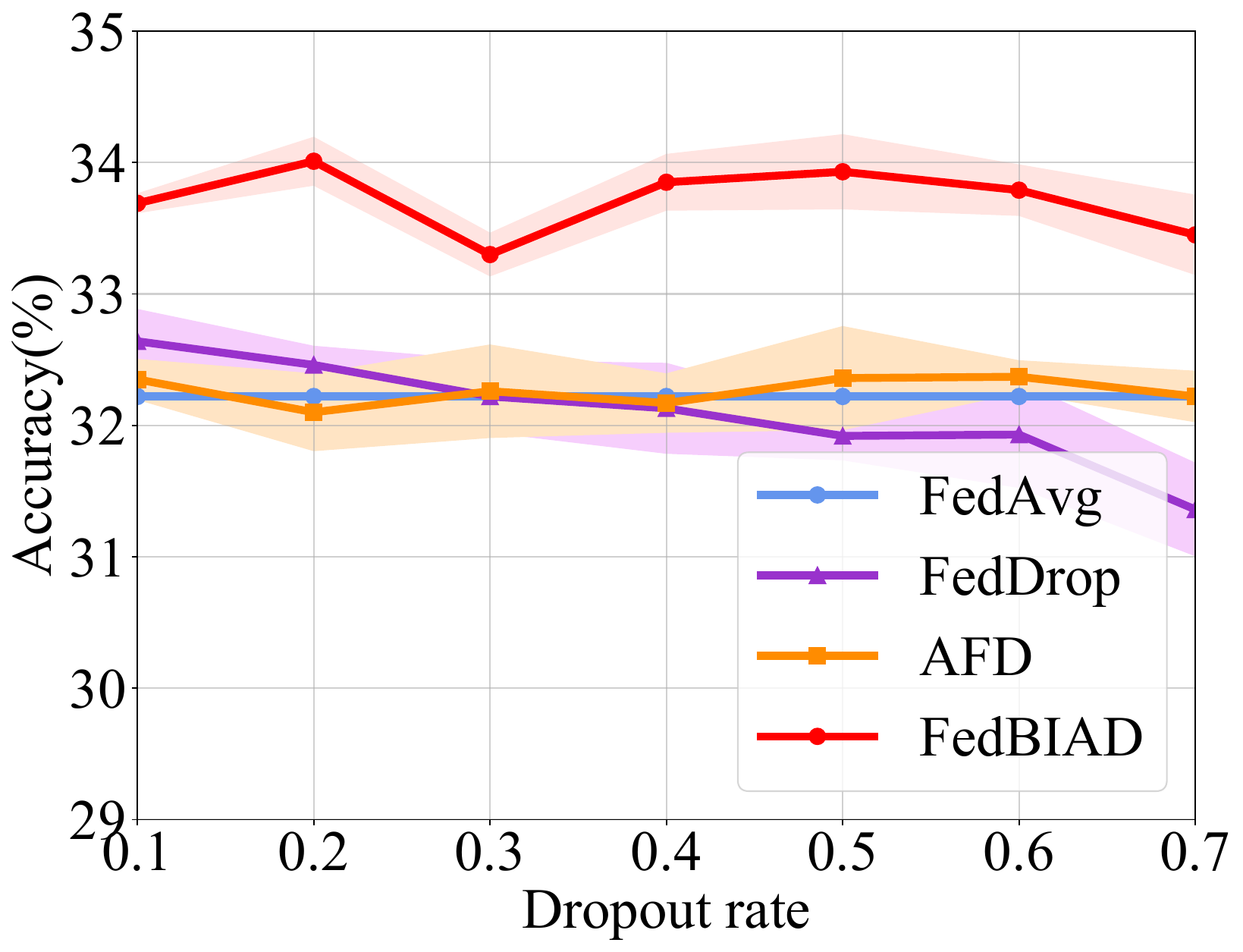}}
				\label{dropout_acc}
			\end{minipage}%
		}%
		\subfloat[TTA versus dropout rate.]{
			\begin{minipage}[t]{0.49\linewidth}
				\centering
				\centerline{\includegraphics[width=0.9\columnwidth]{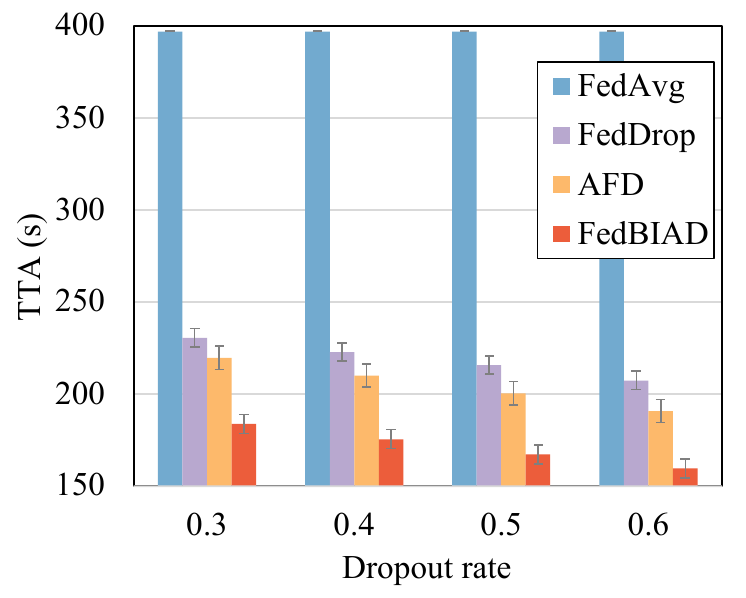}}
				\label{dropout_TTA}
			\end{minipage}
		}
		\centering
		\caption{The results for different dropout rates on Reddit.}
		\label{diff_drop_rate}
		\vskip -0.1in
	\end{figure}
	
	To investigate the effect of dropout rate $p$, we measure accuracy and TTA for different dropout rates on Reddit. As shown in Fig \ref{dropout_acc}, the results of FedAvg without dropout remain constant as it is not affected by dropout rates. We notice that compared with the other three baselines, FedBIAD provides the highest accuracy at different dropout rates. As shown in Fig. \ref{dropout_TTA}, the TTA decreases as the dropout rate increases, and the TTA of FedBIAD is always shorter than other baselines. It indicates that FedBIAD achieves the goal of efficient communication and guaranteed performance in FL.
	
	\section{Conclusion}
	
	In this paper, we propose federated learning with Bayesian inference-based adaptive dropout (FedBIAD) to improve communication efficiency while guaranteeing accuracy. In FedBIAD, we introduce Bayesian inference and adaptively drop weight rows of local models based on the training loss to mitigate uplink overhead. Each client also maintains an importance indicator contributing to finding a high-quality dropping pattern for performance improvement. Theoretical analysis demonstrates that the convergence rate of generalization error of FedBIAD is minimax optimal up to a squared logarithmic factor. Experimental results show that FedBIAD provides up to 2$\times$ reduction in uplink costs with an accuracy increase of up to 2.41\%, which shortens 72\% training time compared to status quo approaches. FedBIAD focuses enough on communication efficiency, and we intend to further explore saving local computation in future work. Another future direction involves the adaptive adjustment of dropout rates based on the changes in communication conditions and device capabilities.
	
	\section*{Acknowledgements}
	This work was supported by the National Natural Science Foundation of China (No. 62072436) and the National Key Research and Development Program of China (2021YFB2900102). 
	
	\appendix
	
	\subsection{Proof of Theorem \ref{thm:upper_bound}}
	
	We want to analyze the upper bound on the generalization error of global variational approximation $\Tilde{\pi}^g$ (see Definition \ref{def:def1} in the paper). Based on Theorem 2 in \cite{Convergence_Bayes_2020}, we can obtain generalization properties of a fully connected DNN, where for any client $k$, the global model $f_\theta$ satisfies
	\begin{align}
		\label{FCNN_bound}
		\mathbb{E} \bigg[ & \|{f_{\theta} -f^k_0}\|^2_2 \Tilde{\pi}^g(\mathrm{d}\theta) \bigg] \notag\\
		& \leq \frac{2}{(1-\alpha)} \inf_{\theta^{\ast}\in \Theta_{S,L,D}}\|f_{\theta^{\ast}} - f^k_0\|_\infty^2  \notag \\
		& + \frac{2}{(1-\alpha)}(1+ \frac{\sigma^2}{\alpha})\varepsilon_{m}^{S,L,D} 
	\end{align}
	with
	\begin{align}
		\label{varepsilon_m_FCNN}
		\varepsilon_{m}^{S,L,D} = &  \frac{SL}{m}\log(2BD) + \frac{S}{4m}\bigg(12\log(LD) + B^2\bigg)  \notag \\
		& + \frac{S}{m} \log\bigg(11d\max(\frac{m}{S},1)\bigg).
	\end{align} In the proof process, \citet{Convergence_Bayes_2020} introduce the optimal posterior variance $\Tilde{s}_m^2 = \frac{S}{16m} \log(3D)^{-1} (2BD)^{-2L} \{(d+1+\frac{1}{BD-1})^2 + \frac{1}{(2BD)^2-1} + \frac{2}{(2BD-1)^2}\}^{-1}$.
	
	Then, we derive the upper bound on the global generalization error of a sparse single-layer RNN that is detailed in Section \ref{formulations}. We start with the variant inequality of theorem 2.6 in \cite{tempered_posteriors_2019}. Given the variational set $\mathcal{F}_{S,T,D}$ of  tempered posterior, if there is $q^{\ast} \in \mathcal{F}_{S,T,D}$ and $\varepsilon_m > 0$ such that
	\begin{equation}
		KL(q^{\ast}\|\Tilde{\pi}^g) \leq m \varepsilon_m
		\label{condition1}
	\end{equation}
	and
	\begin{equation}
		\int \|f_\theta - f_{\theta^{\ast}}\|^2_2 q^{\ast}(d\theta) \leq \varepsilon_m,
		\label{condition2}
	\end{equation}
	then for any $\alpha\in(0,1)$, we get the upper bound on the generalization error of Bayesian variational inference:
	\begin{align}
		\mathbb{E} & \Big[ \int \|{f_{\theta}-f_0^k}\|^2_2 \Tilde{\pi}_{m,\alpha}(d\theta) \Big]  \\ 
		& \leq \frac{2}{1-\alpha} \big( \inf_{\theta^*\in \Theta_{S,T,D}}\|f_{\theta^{\ast}} - f_0^k\|_2^2 + \varepsilon_m \big) + \frac{2\sigma^2\varepsilon_m}{\alpha(1-\alpha)}. \notag
	\end{align} We can prove that our RNN model satisfies Inequality (\ref{condition1}) and (\ref{condition2}) with $q^*(\theta)$ following the distribution of
	\begin{equation*}
		\tilde{\pi}^*_{m,\alpha}(\theta_n) = \beta_n^* \mathcal{N}(\theta^*_n,\Tilde{s}^2) + (1-\beta_n^*)\delta (0) \hspace{0.3cm} \textnormal{for} \hspace{0.2cm} n=1,...,N,
	\end{equation*}
	and
	\begin{align}
		\label{varepsilon_m_RNN}
		\varepsilon_m & = \frac{SL}{m}\log(2BD) + \frac{S}{m}\log(2D^2) \notag \\ 
		& \qquad + \frac{S}{2m} \log\log (3D) + \frac{S B^2}{2m} + \frac{S}{2m} \log\big(\frac{69md^2}{S} \big) \notag \\
		& \leq \frac{SL}{m}\log(2BD) + \frac{SB^2}{2m}\notag \\
		& \qquad + \frac{5S}{2m}\log(2D) + \frac{S}{m} \log\big(9d\max(\frac{m}{S},1)\big)
	\end{align} Here, $\beta_n^*$ is the optimal dropping label of the $n$-th parameter $\theta_n$, and the posterior variance $\Tilde{s}^2$ satisfies
	\begin{align}
		\label{our_variance}
		\Tilde{s}^2 & = \frac{S}{16m d^2} \log(3D)^{-1} (2BD)^{-2L} \\ 
		& \bigg\{ \Big(d+1+\frac{1}{BD-1} \Big)^2 + \frac{1}{(BD)^2-1} + \frac{2}{(BD-1)^2} \bigg\}^{-1}.\notag
	\end{align}Our posterior variance $\Tilde{s}^2$ calculated by (\ref{our_variance}) is larger than $\Tilde{s}_m^2$ used in \cite{Convergence_Bayes_2020}. Thus, (\ref{FCNN_bound}) still holds with $\Tilde{s}^2$ calculated by (\ref{our_variance}).
	
	Consequently, we integrate the derivations on the fully connected DNN and single-layer RNN and extend them to our federated architecture. Combining (\ref{varepsilon_m_FCNN}) and \ref{varepsilon_m_RNN}, we obtain \begin{align}
		\label{varepsilon_m_FedBIAD}
		\varepsilon_{m_r}^{S,L,D} = &  \frac{SL}{m_r}\log(2BD) + \frac{3S}{m_r}\log(LD) + \frac{SB^2}{2m_r} \notag \\ 
		& + \frac{2S}{m_r} \log\Big(4d\max(\frac{m_r}{S},1)\Big),
	\end{align}
	where we consider the different numbers of local data in different clients such that the minimum amount of client-side total input data up to round $r$ is denoted as $m_r=r\times V \times \min\{|\mathcal{D}^1|, \ldots, |\mathcal{D}^K|\}$. The $\varepsilon_{m_r}^{S,L,D}$ denoted as (\ref{varepsilon_m_FedBIAD}) is larger than (\ref{varepsilon_m_FCNN}) and (\ref{varepsilon_m_RNN}) with the same number of input data. Based on (\ref{varepsilon_m_FedBIAD}), we can derive the average generalization error
	of global model $f_{\theta}$ learned by FedBIAD in the server:
	\begin{align}
		\mathbb{E} & \bigg[ \frac{1}{K} \sum_{k=1}^{K} \int \|{f_{\theta}-f^k_0}\|^2_2 \Tilde{\pi}^g(\mathrm{d}\theta) \bigg] \notag\\
		& \leq  \frac{2}{K(1-\alpha)} \sum_{k=1}^{K} \inf_{\theta^{\ast}\in \Theta_{S,L,D}}\|f_{\theta^{\ast}} - f^k_0\|_\infty^2 \notag \\
		& \qquad + \frac{2\sigma^2}{\alpha(1-\alpha)}(1+ \frac{\alpha}{\sigma^2})\varepsilon_{m_r}^{S,L,D}.
	\end{align}
	Therefore, we prove Theorem \ref{thm:upper_bound}.
	
	{\footnotesize
		\bibliographystyle{IEEEtran}
		\bibliography{FedBIAD}
	}
	
\end{document}